\documentclass[letterpaper, amsfonts, amssymb, amsmath, preprint, showkeys, nofootinbib,onecolumn]{revtex4-2}
\usepackage[colorinlistoftodos, color=green!40, prependcaption]{todonotes}
\usepackage[pdftex, pdftitle={Article}, pdfauthor={Author}]{hyperref} 

\usepackage[english]{babel}
\usepackage[utf8]{inputenc}
\usepackage[colorinlistoftodos, color=green!40, prependcaption]{todonotes}

\usepackage{amsthm}
\usepackage{mathtools}
\usepackage{physics}
\usepackage{xcolor}
\usepackage{graphicx}
\usepackage[left=23mm,right=13mm,top=35mm,columnsep=15pt]{geometry}
\usepackage{adjustbox}
\usepackage{placeins}
\usepackage[T1]{fontenc}
\usepackage{lipsum}
\usepackage{csquotes}

\usepackage{siunitx}
\usepackage{rotating}

\usepackage{dcolumn}
\newcolumntype{d}[1]{D{.}{.}{#1}}

\newcommand{\ea}{{\it et al.}}
\begin{document}

\title{Atomistic Simulation Framework for Molten Salt Vapor-Liquid Equilibrium Prediction and its Application to NaCl}

\author{Leann Tran}
        \affiliation{Dept. of Mathematics and Statistics, University of Guelph, Guelph ON N1G 2W1, Canada}
\author{Jiri Skvara}
        \affiliation{Institute of Chemical Process Fundamentals, Czech Academy of Sciences, Prague 6 Suchdol, Czech Republic}
\author{William R. Smith}
        \email[Correspondence email address: ]{bilsmith@uoguelph.ca}
        \affiliation{Dept. of Mathematics and Statistics, University of Guelph, Guelph ON N1G 2W1, Canada}
        \affiliation{Dept. of Chemical Engineering, University of Waterloo, Waterloo ON N2L 3G1, Canada}
        \affiliation{Faculty of Science, Ontario Tech University, Oshawa ON L1H 7K4, Canada}

\date{\today} 

\begin{abstract}
Knowledge of the vapor-liquid equilibrium (VLE) properties of molten salts is important in the design of thermal energy storage systems for solar power and nuclear energy production applications.  The high temperatures involved make their experimental determination problematic, and the development of both macroscopic thermodynamic correlations and  predictive molecular-based methodologies are complicated by the requirement to appropriately incorporate the chemically reacting vapor-phase species.  We derive a general thermodynamic-based atomistic simulation framework for molten salt VLE prediction and show its application to NaCl. Its input quantities are temperature-dependent ideal-gas free energy data for the vapor phase reactions, and density and residual chemical potential data for the liquid.   If these are not available experimentally, the former may be predicted using standard electronic structure software, and the latter by means of classical atomistic simulation methodology.  The framework predicts the temperature dependence of vapor pressure, coexisting phase densities, vapor phase composition, and vaporization enthalpy.  It also predicts the concentrations of vapor phase species present in minor amounts (such as the free ions), quantities that are extremely difficult to measure experimentally.  We furthermore use the VLE results to obtain approximations to the complete VLE binodal dome and the critical properties.   We verify the framework for molten NaCl, for which experimentally based density and chemical potential data are available in the literature.  We then apply it to the analysis of NaCl simulation data for two commonly used atomistic force fields.  The framework can be readily extended to molten salt mixtures and to  ionic liquids.

\end{abstract}


\maketitle

\clearpage
\section{Introduction}\label{sec:1}
The vapor-liquid equilibrium (VLE) properties of molten salts are important in nuclear energy production\cite{LeBrun2007,Benes2009,Ladkany2018,Li2021} and solar power systems\cite{Wang2021,Zhao2021,Caraballo2021}.    Due to the high temperatures involved, the experimental determination of these properties is problematic, and most studies for even the simplest case of pure molten alkali halides date prior to 1980 ({\it e.g.}, Janz \ea\cite{Janz1979}).

Atomistic molecular simulation is a promising alternative approach for molten salt property prediction, but the vast majority of such studies to date have focussed on density and transport properties (for recent reviews, see Wang \ea\cite{Wang2020} and Sharma \ea\cite{Sharma2021}), and its use for VLE prediction has been limited.  In a 1992 study, Panagiotopoulos\cite{Panagiotopoulos1992} applied the Gibbs Ensemble Monte Carlo algorithm \cite{Panagiotopoulos1987,Panagiotopoulos1988}  to the restricted primitive model (RPM) force field (FF)  with its diameter fitted to the NaCl liquid density.  Guissani and Guillot\cite{Guissani1994} in 1994 performed molecular dynamics (MD) simulations in the $NVE$ ensemble ($N$ is the number of particles, $V$ the volume, and $E$ the internal energy)  using the more realistic Born-Huggins-Mayer-Fumi-Tosi (BHMFT) FF\cite{Fumi1964,Tosi1964} with parameters previously reported by Lewis and Singer\cite{Lewis1975}.  They calculated the VLE binodal curve from an empirical equation of state fitted to simulation data for sets of $Pv$ isotherms exhibiting van-der-Waals loops ($P$ is the pressure and $v$ is the molar volume).  The same FF and variants thereof were studied in 2007 by  Rodrigues and Fernandes\cite{Rodrigues2007}(RF), who calculated chemical potentials by the integration of $NVT$ MD simulation isotherms and determined the VLE binodal curve from the equality of the simulated chemical potentials and pressures of the coexisting phases.  Abramo \ea\cite{Abramo2018} recently used a correlation approach to fit BHMFT temperature-dependent ionic size parameters to MD $NVT$ isothermal compressibility simulations, and used these to calculate the resulting VLE envelope using the same methodology as that of RF.  Most recently,  Kussainova \ea\cite{Kussainova2020} (KMYYP) performed MD simulations of NaCl vapor pressures and coexisting phase densities using  two modern electrolyte force fields: that of Joung and Cheatham\cite{Joung2008}, and the Madrid FF of Benavides \ea\cite{Benavides2017}.  They used a treatment that combined $NVT$ liquid chemical potential simulations in conjunction with simulations for a vapor phase model of an ideal gas of ion pairs.

A main challenge for VLE molten salt modeling is the appropriate treatment of the vapor phase (see, for example, Zhao \ea\cite{Zhao2021}), which requires knowledge of the species present. For NaCl and other alkali halides, the vapor phase is experimentally known  to be dominated by the neutral ion-pair monomer and its dimer\cite{Barton1956,Miller1956,Datz1961,Kvande1979,Kvande1979b,Pitzer1996}.  Since molten salt vapor pressures are very low, the vapor phase may be treated as an ideal gas, and its composition is determined by the relevant chemical reaction equilibria, which must be taken into account in the VLE calculations. These were not considered in any of the aforementioned simulation  studies.

The purpose of this paper is to provide a predictive thermodynamic-atomistic simulation framework for  molten salt VLE properties, whose novelty is the integration of ideal-gas chemical reaction free energy quantities with atomistic liquid-state chemical potential and density atomistic simulation data.   This data is required as functions of temperature at the standard state pressure $P^0=1$ bar. For many molten salts ({\it e.g.}, alkali halides), the ideal-gas data are readily available in the NIST-JANAF\cite{Chase1998} or other thermochemical compilations.  If  it is unavailable for a salt of interest, it can be calculated by means of electronic structure methodology such as that incorporated in software such as Gaussian\cite{Gaussian16}.

The  framework also predicts the VLE vaporization enthalpy as a function of temperature, $T$, in addition to the VLE binodal curve and critical properties. The enthalpy has not been considered in prior simulation studies, and its experimental values are sparsely available in the literature (for example, it is not included in the comprehensive review of molten salt properties by Janz \ea\cite{Janz1979}).  Finally, the framework also predicts the concentrations of vapor phase species that are present in minor amounts in a computationally efficient manner.

The paper is organised as follows.  The next section describes the thermodynamic framework.  In the subsequent Results and Discussion section, we first validate the framework using experimental NaCl density data\cite{Kirshenbaum1962} and chemical potential data from the NIST-JANAF compilation\cite{Chase1998} to calculate the vapor pressure as a function of  temperature.  We then show the framework's application to the NaCl FFs previously considered by Kussainova \ea\cite{Kussainova2020}, using their liquid density and chemical potential simulation data.  We calculate the NaCl vapor pressure, the vaporization enthalpy and the VLE phase diagram and critical properties. This is followed by a section giving an uncertainty analysis of our calculation procedures, and then a final  Conclusions section.
\section{Molten Salt VLE Thermodynamic Simulation Framework} \label{sec:framework}
For simplicity and generality, we consider a 1-1 electrolyte AB comprised of an A$^+$ cation and a B$^-$ anion.
\subsection{Vapor Pressure, $P^*$}\label{sec:framework1}
Treating the liquid as an incompressible fluid between $P^0=1$  bar and $P^*$ (the experimental isothermal compressibility of NaCl(liq) is only 0.343 per GPa\cite{Marcus2013}), its mean ionic chemical potential at  $(T,P)$ can be written as
\begin{equation}
\mu_{\rm AB, liq}(T,P)
 = \mu^0_{\rm A^+(IG)}(T,P^0)+\mu^0_{\rm B^-(IG)}(T,P^0)
 +\mu_{\rm AB,liq}^{{\rm res};NPT}(T,P^0)+ \left(\frac{P^0}{\rho_{\rm AB,liq}(T,P^0)RT}\right)\left(\frac{P}{P^0}-1\right)\label{eq:muliqNPT}
\end{equation}
 where $\mu_i^0(T,P^0)$ is the molar standard chemical potential of species $i$ at $T$ and $P^0$, $R$ is the universal gas constant, IG denotes the ideal gas standard state,  and $\mu^{{\rm res};NPT}_{\rm AB,liq}$ is the liquid mean ionic residual chemical potential with respect to the ideal-gas model in the $NPT$ ensemble.

The vapor phase may consist of multiple species, as identified from experimental knowledge or from electronic structure calculations.  Based on experimental information for NaCl and other alkali halides \cite{Barton1956,Miller1956,Datz1961,Kvande1979,Kvande1979b,Pitzer1996}, we begin by assuming only the AB monomer and the (AB)$_2$ dimer species in the vapor phase.  We will show how to verify (or negate) this assumption by considering  additional species after the initial $P^*(T)$ calculation.

At the low pressures typical of molten salt VLE, the  vapor phase species chemical potentials can be modelled by their ideal-gas form:
\begin{equation}
\mu_i(T,P) = \mu^0_i(T,P^0) + RT\ln\left(\frac{y_iP}{P^0}\right) \label{eq:muvap}
\end{equation}
where $y_i$ is the mole fraction of species $i$.
The VLE condition for the vapor pressure $P^*(T)$  arising from the equality of the AB liquid and vapor phase chemical potentials is then
\begin{equation}
\left(\frac{\Delta G_{r1}^0(T,P^0)+\mu^{{\rm res},NPT}_{\rm AB,liq}(T,P^0)}{RT}\right)
 + \left(\frac{P^0}{\rho_{\rm AB, liq}(T,P^0)RT}\right)\left(\frac{P^*}{P^0}-1\right)
 =\ln \left(\frac{y_{\rm AB,vap}P^*}{P^0}\right) \label{eq:Pstarmonomer2}
\end{equation}
where
\begin{eqnarray}
\Delta G_{r1}^0(T,P^0) & = & \mu^0_{\rm A^+(IG)}(T,P^0) + \mu^0_{\rm B^-(IG)}(T,P^0) -\mu^0_{\rm AB}(T,P^0)
\label{eq:DeltaGr}
\end{eqnarray}

The vapor-phase monomer mole fraction $y(T,P^*)$ (henceforth dropping its subscript for notational convenience)  is constrained by the equilibrium condition for its dimerization reaction:
\begin{equation}
(1-y)\exp\left(\frac{\Delta G^0_{r2}(T,P^0)}{RT}\right)- \left(\frac{P^*}{P^0}\right)y^2 =  0 \label{eq:Pfinal2}
\end{equation}
where
\begin{equation}
\Delta G^0_{\rm r2}(T,P^0) = \mu^0_{\rm (AB)_2}(T,P^0)- 2\mu^0_{\rm AB}(T,P^0) \label{eq:Pfinal3}
\end{equation}
$P^*(T)$ is determined from the numerical solution of Eqs. (\ref{eq:Pstarmonomer2}) and (\ref{eq:Pfinal2}).

Pitzer\cite{Pitzer1996} has argued that in addition to the ring form of the dimer in the vapor phase, inclusion of its linear isomer (NaCl)$_2$  is needed at higher temperatures.  Rather than including both species individually in the VLE calculation, they can be incorporated in the analysis  without changing the form of Eq. (\ref{eq:Pfinal2})  by the ``lumping technique" described in Section 9.7 of Smith and Missen\cite{Smith1991b}, which expresses the single (lumped) standard chemical potential of the two $({\rm AB})_2$ isomers as
\begin{eqnarray}
 \mu_{\rm (AB)_2,ring+linear}(T,P) & = &  \mu^0_{\rm (AB)_2,ring+linear}(T,P^0) + RT\ln \left(\frac{y_{\rm (AB)_2,ring+linear} P}{P^0}\right)\nonumber \\
    \mu^0_{\rm (AB)_2,ring+linear}(T,P^0) & = & -RT\ln\left[ \exp\left(\frac{-\mu^0_{\rm (AB)_2, ring}(T,P^0)}{RT}\right)+\exp\left(\frac{-\mu^0_{\rm (AB)_2, linear}(T,P^0)}{RT}\right)\right]\nonumber \\\label{eq:lump}
\end{eqnarray}

The molar fraction $z$ of the ring dimer within the two-member dimer group is given by\cite{Smith1991b}
\begin{equation}
z = \frac{\exp[-\mu^0_{\rm (AB)_2, ring}(T,P^0)/RT]}{\exp[-\mu^0_{\rm (AB)_2, ring}(T,P ^0)/RT] + \exp[-\mu^0_{\rm (AB)_2, linear}(T,P^0)/RT]} \label{eq:fracring}
\end{equation}
For future reference, the lumped dimer ideal-gas species standard enthalpy arising from Eq. (\ref{eq:lump}) is
\begin{eqnarray}
h^0_{\rm (AB)_2, ring+linear}(T,P^0) & = & -RT^2\left(\frac{\partial \mu^0_{\rm (AB)_2, ring+linear}/T}{\partial T}\right)\nonumber \\
& = & zh^0_{\rm (AB)_2, ring }(T,P^0) + (1-z)h^0_{\rm (AB)_2, linear}
\end{eqnarray}

As previously mentioned, we can calculate  the compositions of additional vapor-phase species by first calculating $P^*$ assuming that the monomer and dimers are the only species present, and subsequently performing a vapor-phase reaction equilibrium calculation at the resulting $(T,P^*)$ including the additional species ({\it e.g.}, ${\rm A(g), A_2(g), B(g), B_2(g),  A^+(g), B^-(g)}$) (this requires knowledge of their standard chemical potential data).  This strategy can both serve to verify the neglect of the minor species in the original $P^*$ calculation, and provide an approximation to their concentrations.  The calculation may be rapidly performed using any readily available Gibbs energy minimization algorithm\cite{Smith1991b,Leal2017}. (Iif the monomer and dimer compositions change significantly from the original values in the post hoc calculation, then the original $P^*$ calculation  must be modified to incorporate the additional species at the outset.)

Finally, we note in passing that Eq. (\ref{eq:Pstarmonomer2}) can be written using the identity
\begin{equation}
\mu^{{\rm res},NPT}_{\rm AB,liq}(T,P^0) = \mu^{{\rm res},NVT}_{\rm AB,liq}(T,P^0)- 2RT\ln
\left(\frac{P^0}{\rho_{\rm AB,liq}(T,P^0)RT}\right)\label{eq:convert}
\end{equation}
to give the alternative form
\begin{eqnarray}
\left(\frac{\Delta G_{r1}^0(T,P^0)+\mu^{{\rm res},NVT}_{\rm AB,liq}(T,P^0)}{RT}\right) +2\ln \left(\frac{\rho_{\rm AB,liq}(T,P^0)RT}{P^0}\right)\nonumber \\
 + \left(\frac{P^0}{\rho_{\rm AB, liq}(T,P^0)RT}\right)\left(\frac{P^*}{P^0}-1\right)
& = & \ln \left(\frac{yP^*}{P^0}\right) \label{eq:Pstarmonomer2x}
\end{eqnarray}

Standard chemical potentials may be expressed in many forms.  Molar standard Gibbs formation energies, $\Delta G_{fi}(T,P^0)$ are commonly used, but in this paper we use  the following expression:
\begin{equation}
\mu_i^0(T,P^0) =  \Delta H_{fi}(298.15,P^0) + T{\rm gef}_i(T,P^0) \label{eq:std}
\end{equation}
where gef$_i(T,P^0)$ is the Gibbs Energy Function, $\Delta H_{fi}(T,P^0)$  is the species standard molar enthalpy of formation and
 \begin{eqnarray}
{\rm gef}_i(T,P^0) & = & \left(\frac{G(T,P^0)-H(298.15,P^0)}{T}\right)_i \label{eq:gefdef}
\end{eqnarray}
$\Delta H^0_{\rm r1}(298.15, P^0)$ and $\Delta H^0_{\rm r2}(298.15, P^0)$ data are given in the Supporting Information (SI).

Ideal-gas species molar enthalpy values are expressed as
\begin{equation}
h_i^0(T,P^0)= \Delta H^0_{fi}(298.15,P^0) + [H_i(T,P^0)-H_i(298.15,P^0)]
\end{equation}
and liquid species enthalpy values $h_i(T,P)$ are expressed as
\begin{equation}
h_i(T,P) = h_i^0(T,P^0) + h_i^{{\rm res},\;NPT}(T,P^0)
\end{equation}
where we have neglected the pressure dependence.
\subsection{Vaporization Enthalpy}\label{sec:framework2}

Using the Gibbs-Helmholtz equation for the chemical potentials, the differential of Eq. (\ref{eq:Pstarmonomer2}) on the VLE saturation curve is
\begin{eqnarray}
 -\left[\frac{\Delta H^0_{r1}(T,P^0)+h^{\rm res,NPT}_{\rm AB,liq}(T,P^0)}{RT^2}\right]dT  +\left(\frac{P^*}{\rho_{\rm AB,liq}(T,P^0)RT}-1\right)\frac{dP^*}{P^*}
 -\frac{dy}{y} & = & 0
 \label{eq:reactvap1}
 \end{eqnarray}
 where $y$ is evaluated at $(T,P^*)$ and
 \begin{eqnarray}
 \Delta H^0_{r1}(T,P^0) & = & h^0_{\rm A+}(T,P^0) + h^0_{\rm B-}(T,P^0) -h^0_{\rm AB}(T,P^0) \label{eq:Hvap1}
 \end{eqnarray}
 Similarly, the differential of the reaction equilibrium condition of Eq. (\ref{eq:Pfinal2}) on the VLE saturation curve is
 \begin{equation}
 \left(\frac{\Delta H^0_{r2}(T,P^0)}{RT^2}\right)dT + \frac{dP^*}{P^*} + \left(\frac{2-y}{1-y}\right)\frac{dy}{y}= 0 \label{eq:reactvap2}
 \end{equation}
 where
 \begin{eqnarray}
 \Delta H^0_{r2}(T,P^0) & = & h_{\rm (AB)_2}^0(T,P^0)-2h_{\rm AB}^0(T,P^0)
 \end{eqnarray}
 Eliminating $dy$ from Eqs. (\ref{eq:reactvap1}) and (\ref{eq:reactvap2}) gives the temperature derivative of the vapor pressure:
 \begin{eqnarray}
 \frac{d \ln P^*(T)}{dT} & = & \frac{-(2-y)\Delta H_{r1}(T,P^0) + (1-y)\Delta H^0_{r2}(T,P^0)}{RT^2\left[1-(2-y)P^*v_{\rm AB,liq}(T,P^0)/(RT)\right]}  \label{eq:Clapeyron} 
  \end{eqnarray}
 where
 \begin{equation}
 \Delta H_{r1}(T,P^0) = \Delta H^0_{r1}(T,P^0)+h^{\rm res,NPT}_{\rm AB,liq}(T,P^0) \label{eq:deltahvap1}
 \end{equation}

Eq. (\ref{eq:Clapeyron}) can be written in the Clapeyron form
\begin{eqnarray}
\frac{dP^*(T)}{dT}& = &\frac{\Delta H_{\rm AB}^{\rm vap}(T)}{T\Delta v}
\end{eqnarray}
where
\begin{equation}
 \Delta v = \frac{RT}{P^*}-(2-y)v_{\rm AB, liq}(T,P^0) \label{eq:deltav}
\end{equation}
or alternatively as
\begin{equation}
\frac{d \ln P^*(T)}{d(1/T)}=-\;\frac{\Delta H_{\rm AB}^{\rm vap}(T)}{R\Delta Z} \label{eq:Hvap2}
\end{equation}
 where  $Z$ is the compressibility factor. $\Delta H_{\rm AB}^{\rm vap}(T) $ is the enthalpy change per mole of per mole of vapor formed (from $(2-y)$ mol liquid), given by
 \begin{eqnarray}
 \Delta H_{\rm AB}^{\rm vap}(T) &  =& \; -(2-y)[\Delta H^0_{r1}(T,P^0)+h^{{\rm res},NPT}(T,P^0)] +(1-y)\Delta H^0_{r2}(T,P^0)
  \label{eq:deltaHvapfinz}
 \end{eqnarray}
\section{Results and Discussion}
\subsection{Validation for NaCl using experimental data}\label{sec:validation}
In this section, we validate our thermodynamic framework using experimental data for the required simulation quantities.  We henceforth refer to this as the ``NJ" (for ``NIST-JANAF") methodology.

$\mu^{{\rm res},NPT}_{\rm NaCl,liq}(T,P^0)$ values  were calculated from Eq. (\ref{eq:muliqNPT}) and species Gibbs Energy Function data in the NIST-JANAF compilation\cite{Chase1998} from
\begin{equation}
\mu_{\rm NaCl,liq}^{{\rm res};NPT}(T,P^0) = \mu^0_{\rm NaCl, liq}(T,P^0) - \mu^0_{\rm Na^+(IG)}(T,P^0)-\mu^0_{\rm Cl^-(IG)}(T,P^0) \label{eq:getmuresNPT}
\end{equation}
We smoothed the  data for use in Eqs. (\ref{eq:Pstarmonomer2})-(\ref{eq:Pfinal3}) by fitting each of  $\Delta G^0_{r1}(T,P^0)/T$, $\Delta G^0_{r2}(T,P^0)/T$ and $\mu^{{\rm res},NPT}(T,P^0)/T$ to the functional form
\begin{equation}
f_1(T) = A + \frac{B}{T} + C\ln T \label{eq:Gfit}
\end{equation}
We used the interval (1200 K, 2500 K) for the NJ $\Delta G^0(T,P^0)$ and $\mu^{{\rm res},\;NPT}(T,P^0)$ data, in conjunction with the density data of Kirschenbaum \ea\cite{Kirshenbaum1962} (see Table \ref{table:PstarK}), which we fitted to the functional form
\begin{equation}
f_2(T) = a+bT+cT^2 \label{eq:rhofit}
\end{equation}
over the interval (1149 K, 3200 K). The regression coefficients and the standard regression errors are given in the SI.

\begin{figure}[h!] 
\includegraphics[width=0.7\textwidth]{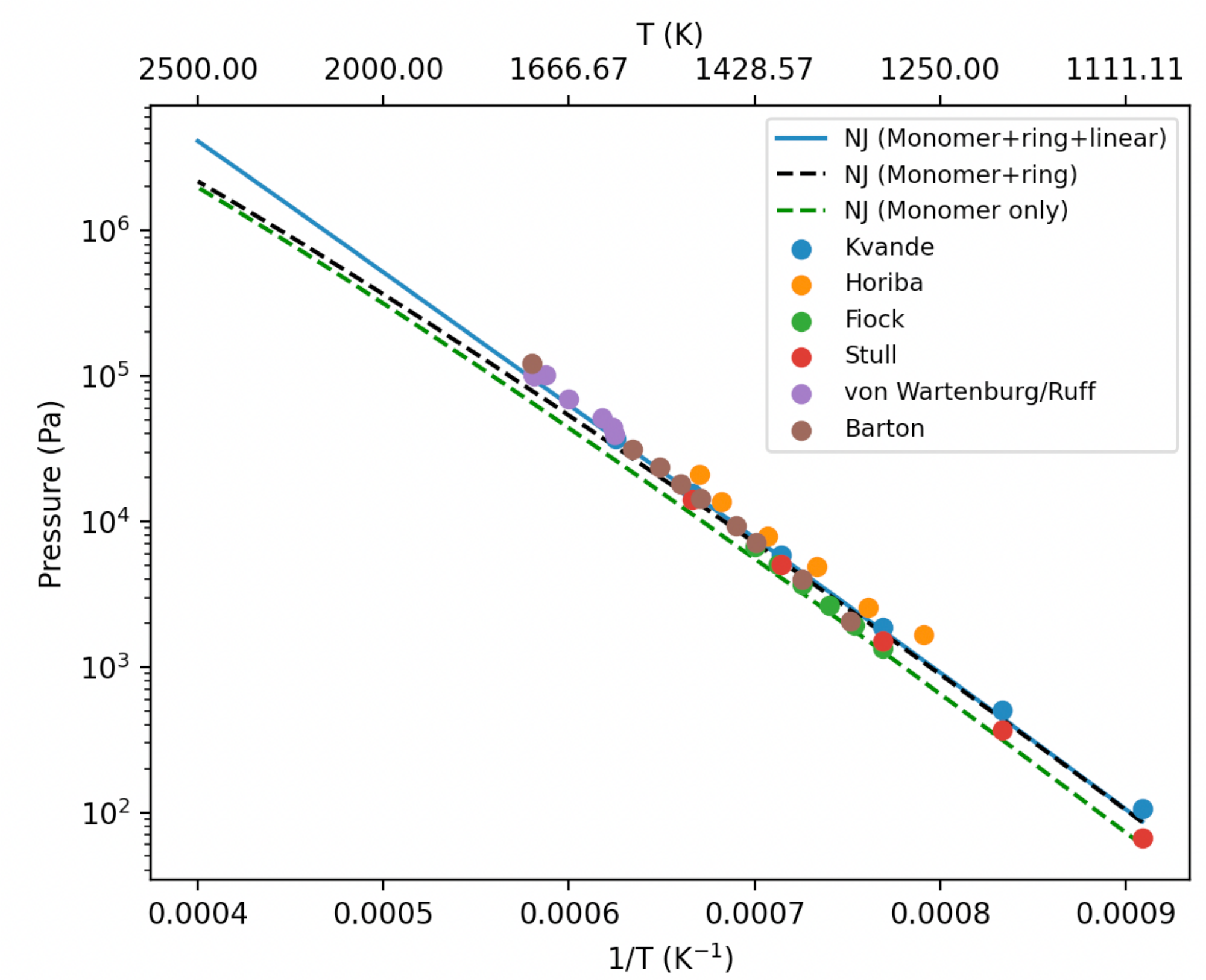}
    \caption{\noindent NaCl vapor pressure as a function of temperature calculated from Eqs. (\ref{eq:Pstarmonomer2})-(\ref{eq:Pfinal3}) using the experimentally based NJ methodology and incorporating different sets of vapor-phase species. The filled circles  indicate  experimental results as follows. blue: Kvande\cite{Kvande1979b}, orange: Horiba\cite{Horiba1928}, green: Fiock\cite{Fiock1926}, red: Stull\cite{Stull1947a}, purple: von Wartenberg \ea\cite{Von1921} and Ruff\cite{Ruff1921}, brown: Barton\cite{Barton1956}.}
    \label{fig:exppressures}
\end{figure}

Fig. \ref{fig:exppressures} illustrates the effects of including different sets of species in the vapor phase using the experimentally based NJ approach. Shown are calculations of the NaCl vapor pressures using Eqs.(\ref{eq:Pstarmonomer2})-(\ref{eq:Pfinal3}) over the temperature range (1100 K, 2500 K) using three different vapor-phase species sets: (1) only the monomer NaCl(g); (2) the monomer and the ring dimer, $({\rm NaCl})_2$; and (3) the monomer and both ring and linear dimers. Numerical data for case (3) is given in Table \ref{table:Pstarvalues}  and for the other cases, it is provided in the SI. At about 2300 K and higher temperatures where the vapor pressure exceeds 10 bar (and no experimental data exist), the vapor phase ideal-gas approximation introduces errors that will increase with temperature.  These could be reduced by incorporating second virial coefficients, but this is beyond the scope of the current study.

\begin{table}[htbp] 
\vspace*{-1cm}
\begin{center}
\begin{tabular}{|ccccrc|}
\hline
$T$    & $P^*$        & $\Delta H_{\rm NaCl}^{\rm vap}$ & $\rho_{\rm vap}$ & \multicolumn{1}{c}{$\rho_{\rm liq}$} & Data Used\\[-8pt]
    (K) & (Pa) & (kJ$\cdot$mol$^{-1}$) &  (kg$\cdot$m$^{-3}$) & \multicolumn{1}{c}{(kg$\cdot$m$^{-3}$)} & \\
\hline
            & 4.45E+02  & 179.5 & 2.61E-03 & 1486.8  & NJ \\[-8pt]
 1200   & 5.57E+03  & 169.4 & 5.51E-02 & 1345.1  & JC \\[-8pt]
            & 3.34E+01  & 221.6 & 2.03E-04 & 1264.7  & MFC-MSC \\
\hline
            &  1.77E+03 & 178.2 & 9.56E-03 & 1440.7 & NJ \\[-8pt]
 1300   &  2.06E+04 & 169.3 & 1.86E-01 & 1294.5 & JC \\[-8pt]
            &  1.86E+02 & 224.8 & 1.06E-03 & 1203.8 & MFC-MSC \\
\hline
            &  5.72E+03 & 176.7 & 2.87E-02  & 1394.2  & NJ \\[-8pt]
 1400   & 6.29E+04  & 169.0 & 5.26E-01  & 1244.2 & JC \\[-8pt]
            & 8.33E+02  & 228.5 & 4.47E-03  & 1140.2 & MFC-MSC \\
\hline
            & 1.57E+04 & 175.6 & 7.35E-02  & 1347.3 & NJ \\[-8pt]
 1500   & 1.65E+05 & 168.6 & 1.29E+00 & 1194.2 & JC \\[-8pt]
            & 3.12E+03 & 233.1 & 1.60E-02 & 1073.7 & MFC-MSC \\
\hline
            & 3.77E+04 &174.5  & 1.66E-01  & 1300.1 & NJ \\[-8pt]
 1600   & 3.85E+05 & 168.3 & 2.83E+00 & 1144.6 & JC \\[-8pt]
            & 1.02E+04 & 238.8 & 5.03E-02  & 1004.4 & MFC-MSC \\
\hline
            & 8.15E+04 & 173.6 & 3.37E-01  & 1252.5 & NJ \\[-8pt]
 1700   & 8.12E+05 & 168.0 & 5.65E+00 & 1095.3 &JC \\[-8pt]
            & 2.97E+04 & 246.1 &1.44E-01   & 932.3  & MFC-MSC \\
\hline
            & 1.61E+05 & 172.9 &  6.29E-01 & 1204.6 & NJ \\[-8pt]
 1800   & 1.58E+06  & 168.0 & 1.05E+01& 1046.3 & JC \\[-8pt]
            & 7.95E+04  & 255.4 & 3.80E-01 & 857.4   & MFC-MSC \\
\hline
            &  2.96E+05 & 172.6 & 1.09E+00 & 1156.2 &NJ \\[-8pt]
 1900   & 2.87E+06  & 168.1 & 1.82E+01 & 997.6  & JC \\[-8pt]
            & 1.99E+05  & 267.2 & 9.52E-01  &  779.7& MFC-MSC \\
\hline
            & 5.11E+05  & 172.5  & 1.80E+00 &1107.5 & NJ \\[-8pt]
 2000   & 4.96E+06  & 168.3 & 3.01E+01 &  949.3 & JC \\[-8pt]
            &  4.76E+05 & 281.5 & 2.29E+00 &  699.2 & MFC-MSC \\
\hline
            & 8.39E+05  & 172.8  &  2.81E+00& 1058.5 & NJ \\[-8pt]
 2100   & 8.21E+06  & 168.8 & 4.79E+01  &  901.3 & JC \\
 \hline
\end{tabular}
\caption{NaCl VLE properties calculated by the framework of Section \ref{sec:framework1} using different liquid data, with an ideal-gas  vapor phase of  monomer plus ring and linear dimers. In all cases, $\Delta G^0_{r1}(T,P^0)$ and $\Delta G^0_{r2}(T,P^0)$ were obtained from the NIST-JANAF compilation\cite{Chase1998}, augmented with the Pitzer\cite{Pitzer1996} linear dimer data.  NJ: Kirshenbaum density data\cite{Kirshenbaum1962} and NIST-JANAF $\mu^{{\rm res}\;NPT}_{\rm NaCl}(T,P^0)$ data;  JC: Joung-Cheatham FF\cite{Joung2008} density and $\mu^{{\rm res}\;NPT}_{\rm NaCl}(T,P^0)$ data; MFC-MSC: MSC density data and the hybrid method for calculating $\mu^{{\rm res}\;NPT}_{\rm NaCl}(T,P^0)$ for the Madrid FF \cite{Benavides2017} (denoted by Kussainova \ea\cite{Kussainova2020} as the ``Madrid rerun").
}
\label{table:Pstarvalues}
\end{center}
\end{table}
Table \ref{table:verify} shows vapor-phase ideal-gas reaction equilibrium calculations that include additional species in the vapor phase at the example temperatures 1200 K and 2100 K and the corresponding NJ $P^*(T)$ values in Table \ref{table:Pstarvalues}.   In addition to providing useful additional information, these indicate that the concentrations of the additional species are very small, and that they thus have an insignificant effect on the originally calculated VLE results.    The atomic Na and Cl species have the next largest concentrations, and the ion concentrations are vanishingly small at 1200 K and larger but still negligible at 2100 K.
\begin{table} 
\begin{center}
\begin{tabular}{|c|c|c|c|c|c|}
\hline
   &\multicolumn{2}{c|}{1200 K, 445 Pa} & \multicolumn{2}{c|}{2100 K, 8.39 bar}\\
\hline
Species & All  & Table \ref{table:Pstarvalues}& All  & Table \ref{table:Pstarvalues} \\
\hline
NaCl	          &	0.712       & 0.712 & 0.591           & 0.592 \\	
(NaCl)$_2$    &	0.288	    & 0.288  & 0.409          & 0.408\\	
Na	              & 	4.08E-06	& $-$     &  6.63E-04    & $-$\\
Cl	                  & 	4.07E-06	& $-$     &  6.57E-04    &$-$\\
Na$_2$	      &	7.80E-15	& $-$     &  1.42E-08    &$-$\\
Cl$_2$	          &	3.31E-09	& $-$     &  3.23E-06    &$-$\\
Na$^+$	      &	8.49E-10	& $-$     &  3.09E-06    &$-$\\
Cl$^-$	          &	8.49E-10	& $-$     &  3.09E-06    &$-$\\
\hline
\end{tabular}
\caption{Equilibrium vapor-phase mole fractions at 1200 K and \mbox{2000 K} and the corresponding NaCl vapor pressures  calculated from the NJ-based framework of Section \ref{sec:framework1} in Table \ref{table:Pstarvalues} with the vapor phase containing the indicated additional vapor-phase species.  The additional ideal-gas thermochemical data were obtained from the NIST-JANAF compilation\cite{Chase1998}.}
\label{table:verify}
\end{center}
\end{table}

Fig. \ref{fig:exppressures} shows that the $P^*(T)$ data from different experimental groups show some scatter, and the distinctions among the calculated curves are small but apparent on the logarithmic scale of the graph. The solid blue NJ (monomer+ring+linear dimer)  curve shows essentially perfect agreement with the experimental data, and will be treated henceforth as the benchmark for comparison with subsequently shown results obtained from simulation data.  The green dashed curve, which includes only the vapor-phase monomer, somewhat under-predicts the experimental data and the NJ curve at all temperatures.  The black dashed curve, including the monomer and the ring dimer, is in good agreement with the experimental data and the NJ results at temperatures below about 1400 K, but deviates increasingly from them at higher temperatures, consistent with the observation of Pitzer\cite{Pitzer1996}.
\subsection{Application to NaCl  molecular simulation data}
In a recent study, Kussainova \ea\cite{Kussainova2020} (KMYYP) performed MD density and $\mu^{{\rm res}\;NVT}$ simulations for NaCl, using the nonpolarizable FF of Joung and Cheatham\cite{Joung2008} (JC) and variants of the Madrid scaled-charged FF of Benavides \ea\cite{Benavides2017} (MSC).  They used this data to apply a thermodynamic framework different from that of Section \ref{sec:framework} to calculate $P^*(T)$.

Finding the MSC liquid chemical potential to significantly overestimate the experimental NIST-JANAF\cite{Chase1998} values, they followed an earlier suggestion of Vega \cite{Vega2015}  and calculated its chemical potential by insertion of a fully charged ion pair into configurations of an MSC molecular dynamics simulation trajectory, which they called the ``Madrid rerun" approach.  We refer to this in the following as the ``MFC-MSC" methodology.

In the next two subsections we briefly review the KMYYP density and chemical potential results and their agreement with experiment, and in the subsequent two subsections we  apply the framework of Section \ref{sec:framework1} to calculate $P^*(T)$ and the vaporization enthalpy $\Delta H^{\rm vap}(T)$ (not calculated in the KMYYP study) using their simulation data in conjunction with ideal-gas thermochemical data from the NIST-JANAF compilation\cite{Chase1998}.
\subsubsection{Density}\label{sec:density}
\begin{figure}[h] 
    \centering
    \includegraphics[width=0.7\textwidth]{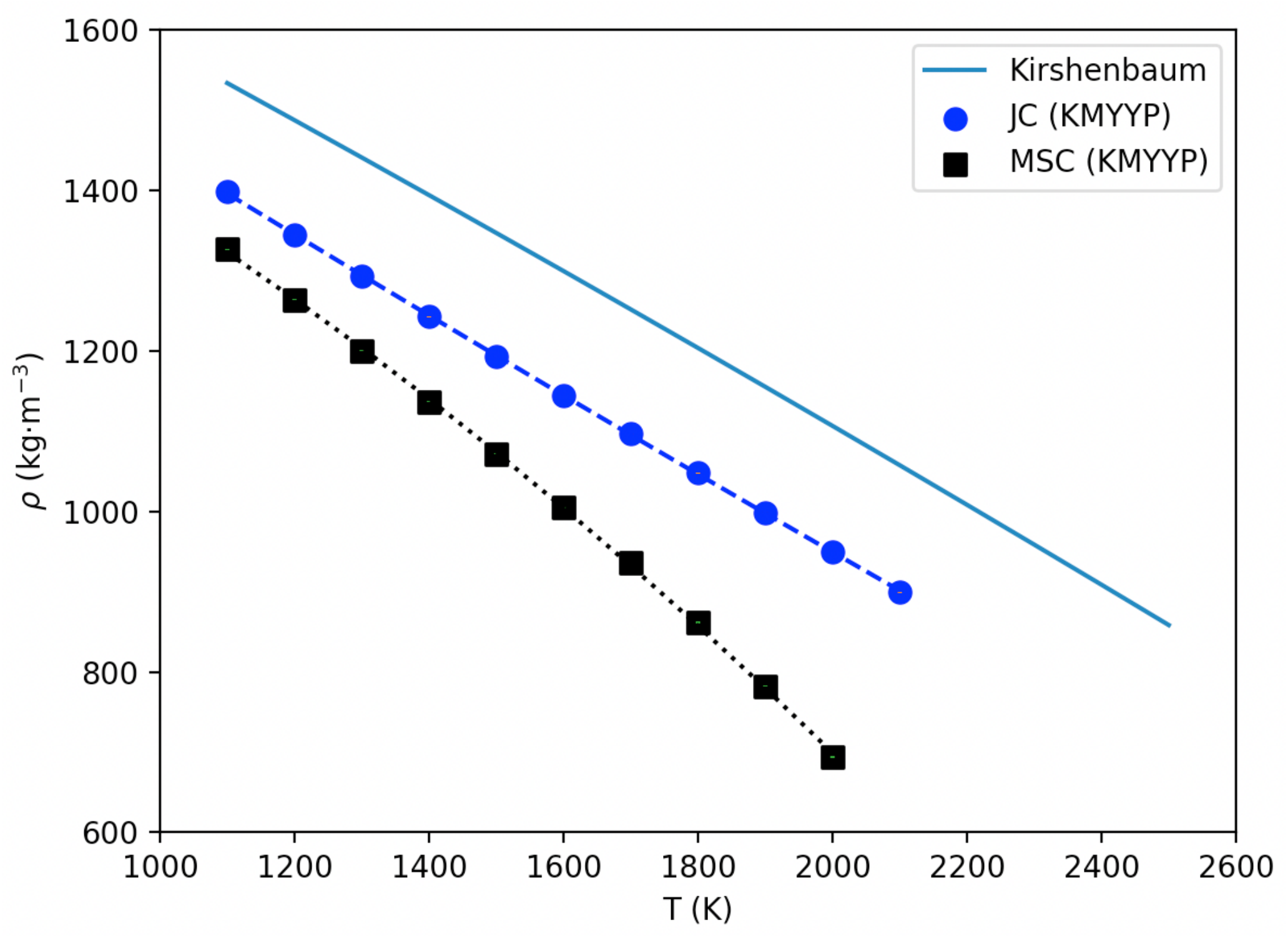}
    \caption{Kussainova \ea\cite{Kussainova2020} (KMYYP) simulation results for the NaCl liquid densities from the Joung-Cheatham\cite{Joung2008}  (JC) and the (scaled-charge) Madrid\cite{Benavides2017} (MSC) force fields, and experimental densities from Kirshenbaum \ea\cite{Kirshenbaum1962} as functions of temperature at $P^0=1$ bar.  The simulation uncertainties lie within the symbols. }
    \label{fig:densities}
\end{figure}

We used the regression form of Eq. (\ref{eq:rhofit}) in Section \ref{sec:validation} to smooth the KMYYP JC and MSC density simulation data\cite{Kussainova2020} over the temperature range for which the data was available (the regression coefficients and standard errors of the fits are given in  the SI).  The smoothed curves and the underlying data are shown in Fig. \ref{fig:densities} in comparison with the experimental data of Kirshenbaum \ea\cite{Kirshenbaum1962} fitted to the same functional form.  Both force fields give somewhat low densities in comparison with experiment, and the Joung-Cheatham\cite{Joung2008}(JC) simulation results give the best agreement.

\subsubsection{Residual chemical potential}\label{sec:mures}
The $\mu^{\rm total}_{\rm NaCl,liq}(T,P^0)$  liquid data  shown in Fig. 2 of Kussainova \ea\cite{Kussainova2020}  and calculated from Eqs. (3) and (4) of their paper use NIST-JANAF standard Gibbs formation free energies for Na$^+$ and Cl$^-$ in conjunction with their density and $\mu_{\rm NaCl,liq}^{{\rm res},NVT}$ simulation data. We obtained  $\mu_{\rm NaCl,liq}^{{\rm res},NPT}(T,P^0)$ simulation values for each FF from  Eq. (\ref{eq:getmuresNPT}) using their total $\mu^{\rm total}_{\rm NaCl,liq}(T,P^0)$ values.

\begin{figure}[h] 
    \centering
    \includegraphics[width=0.7\textwidth]{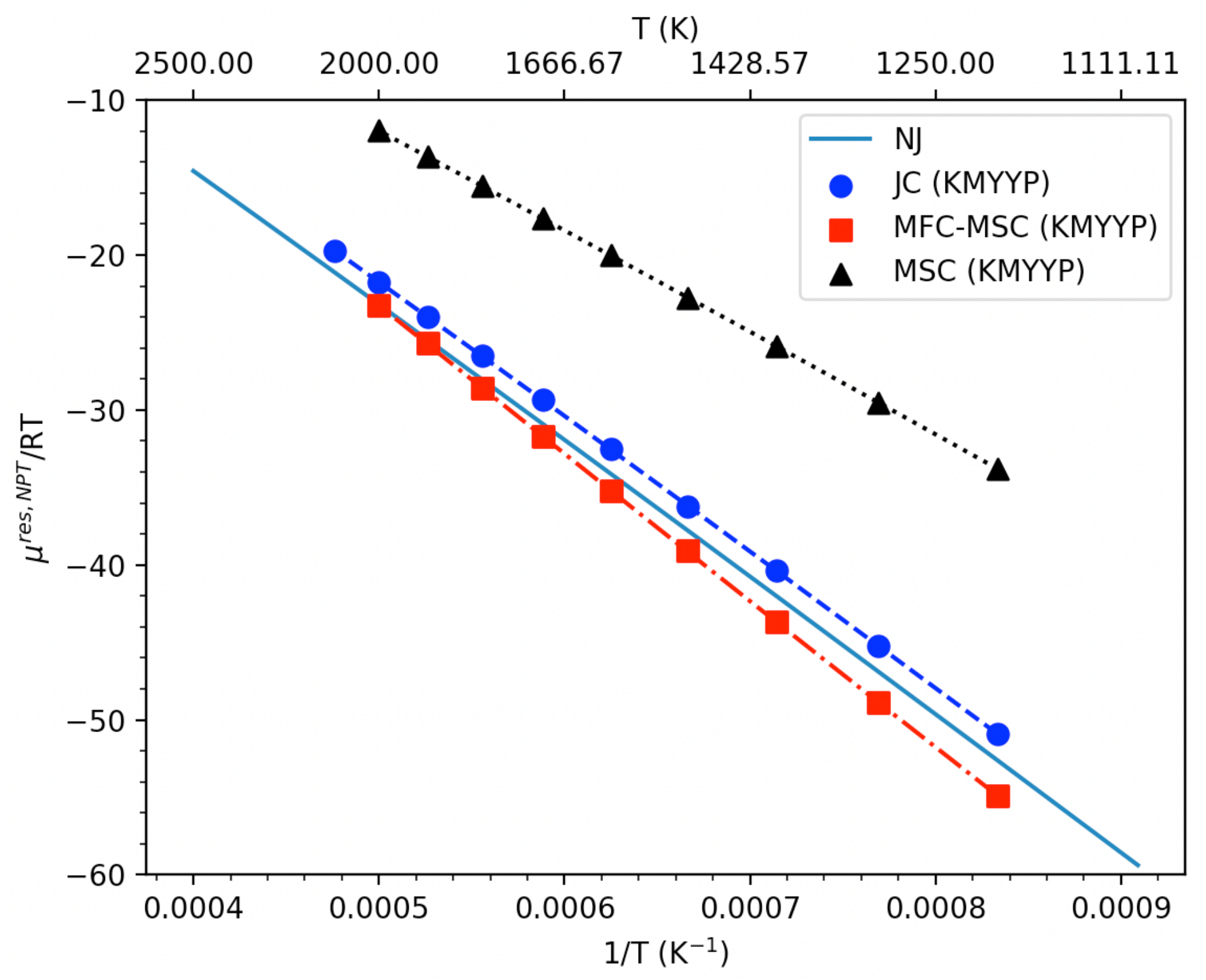}
    \caption{$\mu^{{\rm res},\;NPT}_{\rm NaCl,liq}(T,P^0)/RT$  obtained from Eq. (\ref{eq:getmuresNPT}) as functions of reciprocal absolute temperature.   NJ: regression of the results using NIST-JANAF\cite{Chase1998} $\Delta G_{fi}(T,P^0)$ for NaCl(liq), Na$^+$ and Cl$^-$.  The  curves are the regressions of the points (see the SI), and the simulation uncertainties lie within the symbols.
}
    \label{fig:mures}
\end{figure}

We  smoothed the resulting $\mu^{{\rm res},NPT}_{\rm NaCl,liq}(T,P^0)$ values (see the SI for details) using the functional form of Eq. (\ref{eq:Gfit}). The  results are shown in Fig. \ref{fig:mures} for the JC and for both Madrid force fields (the original scaled charge version and the hybrid MFC-MSC ``Madrid rerun" calculation), where they are compared with the corresponding experimental quantity obtained from the NIST-JANAF compilation\cite{Chase1998}.

The JC values are almost parallel to and over-predict the experimentally based NJ values by 1.4 to 1.7 units, the MFC values under-predict them by 0.1 to  2.3 units, and the MSC values over-predict them by 11-19 units.
\subsubsection{Vapor Pressure}
\begin{figure}[h] 
\centering
\includegraphics[width=0.7\textwidth]{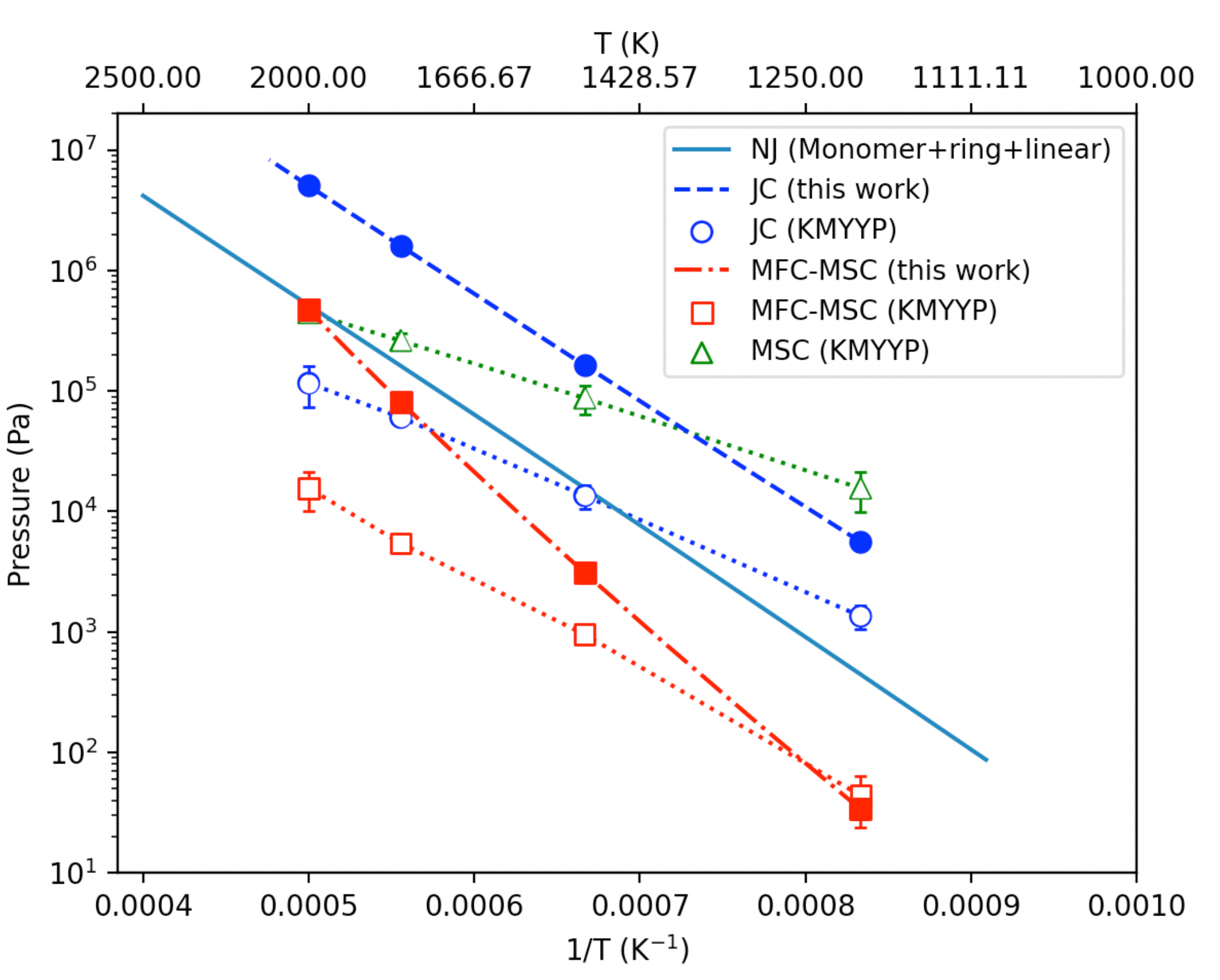}
 \caption{
NaCl vapor pressures as functions of temperature calculated from this work (denoted by ``this work") and from Kussainova \ea\cite{Kussainova2020} (denoted by ``KMYYP") in comparison with the experimentally based NJ results.
The error bars (one standard deviation) on the KMYYP curves are taken from their paper.  The error bars on the other curves were calculated using the methodology of Section \ref{sec:uncertaintyP}, and lie within the symbols.
}.
     \label{fig:ffcomparison}
\end{figure}

Fig. \ref{fig:ffcomparison}  shows vapor pressures (filled symbols joined by curves) calculated from Eqs. (\ref{eq:Pstarmonomer2})-(\ref{eq:Pfinal3}) incorporating the monomer and both dimers in the vapor phase using the KMYYP density and chemical potential data of Figs. \ref{fig:densities} and \ref{fig:mures} in conjunction with the NIST-JANAF\cite{Chase1998} and Pitzer\cite{Pitzer1996} ideal-gas data for $\Delta G^0_{r1}(T,P^0)$ and $\Delta G^0_{r2}(T,P^0)$.  They are compared with the KMYYP results (open symbols joined by dotted curves) shown in Fig. 4 of their paper\cite{Kussainova2020} and with the NJ solid blue curve, replicated from Fig. \ref{fig:exppressures} for comparison.
Our JC results are about an order of magnitude higher than and almost parallel to the NJ curve  (the latter aspect will be seen in the next section to have important implications for the vaporization enthalpy results).  Our MFC-MSC results, indicated by the red dash-dot curve, are lower than the NJ  curve at low temperatures and increase linearly with temperature until they intersect the NJ  curve at about 2000 K.  Although these trends with respect to the  NJ curve are similar to the trends followed by their respective $\mu^{{\rm res}\;NPT}$  results in Fig. \ref{fig:mures}, differences between the curves are magnified  in Fig. \ref{fig:ffcomparison} due to  the logarithmic term involving $P^*$ in Eq. (\ref{eq:Pstarmonomer2}).
The best overall agreement with the NJ curve is achieved by the MFC-MSC approach.   Interestingly, no solutions to Eqs. (\ref{eq:Pstarmonomer2})-(\ref{eq:Pfinal3}) were found to exist in the case of the original scaled-charge Madrid FF (MSC). This is a consequence of the large positive deviation of its $\mu^{{\rm res}\;NPT}$  from the NJ results in Fig. \ref{fig:ffcomparison}.

Our results in Fig. \ref{fig:ffcomparison} using the framework of Section \ref{sec:framework} are seen to differ markedly from the KMYYP\cite{Kussainova2020} results, due to different approaches are used to calculate the ideal-gas $\Delta G^0_{r1}$ quantity, and an incorrect factor of 2 in Eq. (6) of their paper.
\subsubsection{Vaporization enthalpy}
 \begin{figure}[h] 
    \centering
    \includegraphics[width=0.6\textwidth]{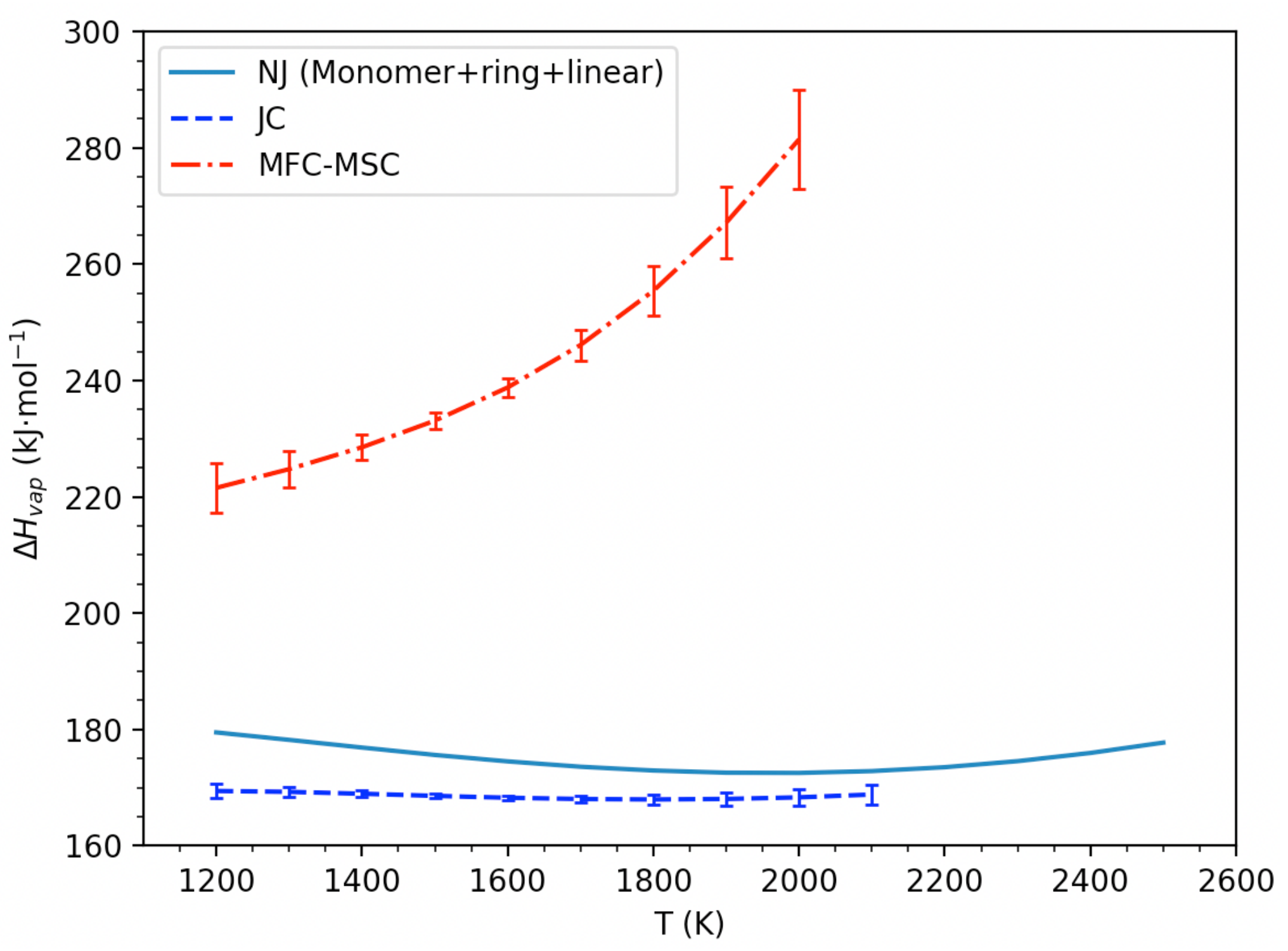}
    \caption{NaCl vaporization enthalpy as a function of temperature calculated from Eq. (\ref{eq:deltaHvapfinz}) using different input data.  The error bars (one standard deviation) were calculated using the methodology of Section \ref{sec:uncertaintyP}}.
    \label{fig:enthalpy}
\end{figure}
We obtained $\Delta H_{\rm NaCl}^{\rm vap}(T)$ from (\ref{eq:deltaHvapfinz}) using the liquid residual enthalpy from differentiation of the regression equation for $\mu^{{\rm res}\;NPT}_{\rm NaCl,liq}$
 in Eq. (\ref{eq:Gfit}):
\begin{eqnarray}
h^{{\rm res},NPT}_{\rm NaCl,liq}(T,P^0) & = & -T^2\left(\frac{\partial \mu^{{\rm res},\;NPT}(T,P^0)/T}{\partial T}\right)\nonumber \\
& = & B -CT \label{eq:deltaHvapfinvapenthalpy}
\end{eqnarray}

Values of $\Delta H_{\rm NaCl}^{\rm vap}(T)$ of. (\ref{eq:deltaHvapfinz}) are given in Table \ref{table:Pstarvalues} and shown in Fig. \ref{fig:enthalpy}.  The  experimentally-based NJ curve shows a weak temperature dependence, and the JC curve is almost flat and lies very close to it. This good agreement is due in large part to the parallelism of the corresponding $P^*(T)$ curves in Fig. \ref{fig:ffcomparison}   The MSC-MFC curve is seen to lie well above the other results.

Experimental data for $\Delta H_{\rm NaCl}^{\rm vap}(T)$ is sparse, and unlike the analysis of Section \ref{sec:framework2}, have been derived indirectly from regressions of $\ln[P^*(T)]$ curves under various assumptions.  Thus, assuming only monomer species in the vapor phase, Fiock and Rodebush\cite{Fiock1926} calculated a constant value of 180 kJ$\cdot$mol$^{-1}$ over the temperature range (1180 K-1429 K); Kelley\cite{Kelley1935} re-analyzed their data and obtained $\Delta H^{\rm vap}_{\rm NaCl}=4.184(52.800-0.0069T)$  kJ$\cdot$mol$^{-1}$, which gives decreasing values of 186-180 kJ$\cdot$mol$^{-1}$ over the same temperature range.  Barton and Bloom\cite{Barton1956} give a value of $180.9\pm 1.7$ kJ$\cdot$mol$^{-1}$
at 1339 K.
\subsection{NaCl VLE Phase Diagram and Critical Properties}
Except near the critical point, the coexisting vapor densities are very small with respect to the liquid densities, and special care must be taken in modeling them simultaneously in order to avoid the prediction of negative vapor densities.  We thus used the method of Rodrigues and Fernandes\cite{Rodrigues2007}, which models the liquid-vapor coexistence curve using the following scaling law equations:
\begin{eqnarray}
\frac{
\sqrt{\rho_{\rm liq}}-\sqrt{\rho_{\rm vap}}
}
{2}& = & E_0\tau^{1/2} + E_1\tau^{3/2} \label{eq:regress1}\\
\frac{\sqrt{\rho_{\rm liq}}+\sqrt{\rho_{\rm vap}}}{2}& = &\sqrt{\rho_c} +F_0\tau+F_1\tau^{3/2} \label{eq:regress2}
\end{eqnarray}
where $T_c$ and $\rho_c$ are respectively the critical temperature and density, $(E_0,E_1,F_0,F_1)$ are parameters, and  $\tau=1-T/T_c$, leading to the following expressions for the densities:
\begin{eqnarray}
\sqrt{\rho_{\rm liq}}   & = & \sqrt{\rho_c} +E_0\tau^{1/2}+F_0\tau+(E_1+F_1)\tau^{3/2}\\
\sqrt{\rho_{\rm vap}} & = & \sqrt{\rho_c}-E_0\tau^{1/2} +F_0\tau+(F_1-E_1)\tau^{3/2}
\end{eqnarray}
Our  analysis proceeds by first fitting the density data to Eq. (\ref{eq:regress1})  to determine $(T_c,E_0,E_1)$, and then using the resulting $T_c$ value to determine $(\rho_c,F_0,F_1)$  by fitting the data to Eq. (\ref{eq:regress2}).  The first regression can be treated as a separable least-squares problem with the nonlinear parameter $T_c$ and the linear parameters $(E_0,E_1)$, and the second regression is linear in the parameters $(F_0,F_1)$.

Following the determination of $T_c$, we fitted the P*(T) data to the Antoine equation to determined the critical pressure, $P_c$.
\begin{equation}
\log_{10}(P) = A-\frac{B}{C+T}
\end{equation}
The Antoine parameter values are given in the SI

\begin{table}[h] 
\begin{center}
\begin{tabular}{|crS|}
\hline
\multicolumn{1}{|c}{$T$}       & \multicolumn{1}{c}{$\rho_{\rm liq}$}            & \multicolumn{1}{c|}{$\rho_{\rm vap}$}\\
\multicolumn{1}{|c}{ (K)}       &  \multicolumn{1}{c}{(kg$\cdot$m$^{-3}$)}   & \multicolumn{1}{c|}{(kg$\cdot$m$^{-3}$)}\\
\hline
1738   & 1233 & 0.577\\
1800   & 1203 & 0.837\\
2000   & 1107 & 2.25\\
2200   & 1009 & 4.95 \\
2400   &  909  & 10.\\
2600   &  807  & 17.4\\
2800   &  701  & 41.\\
3000   &  622  & 59.\\
3200   &  480  & 100. \\
\hline
\end{tabular}
\caption{NaCl vapor-liquid equilibrium density data of Kirshenbaum\cite{Kirshenbaum1962} at temperatures greater than or equal to the NaCl normal boilng point of 1738 K.
}
\label{table:PstarK}
\end{center}
\end{table}
 \begin{figure}[h] 
    \centering
    \includegraphics[width=0.8\textwidth]{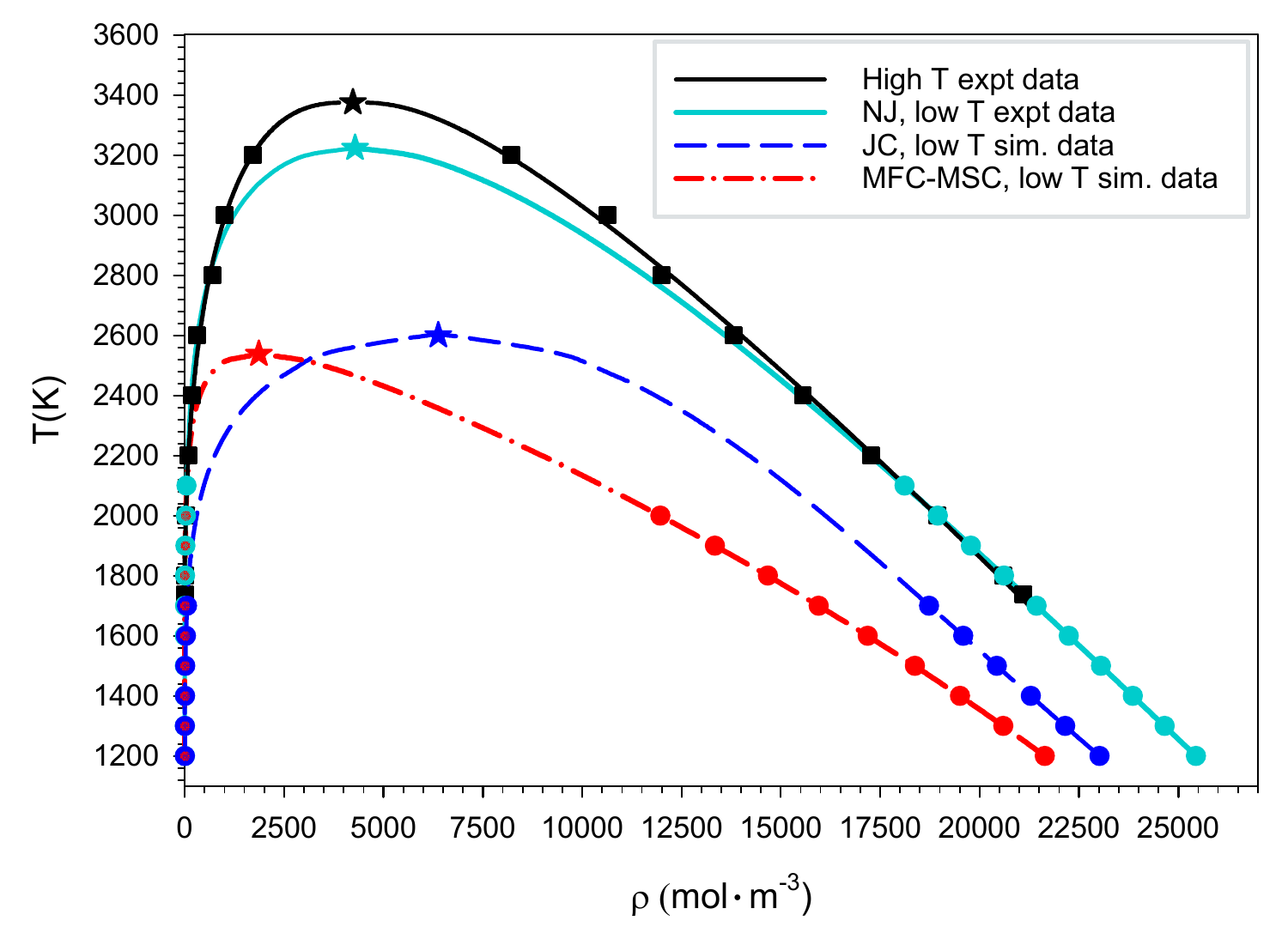}
    \caption{NaCl VLE phase diagram.  Curves are fits to the indicated data points using the scaling law methodology of Rodrigues and Fernandes\cite{Rodrigues2007} and the stars indicate the critical points of the last four rows of Table \ref{table:critical}.
    }
    \label{fig:dome}
\end{figure}

\begin{table}[h!] 
\begin{center}
\begin{tabular}{|c|l|l|c|}
\hline
Data Source/Analysis Method          & \multicolumn{1}{c|}{$T_c$(K) } & \multicolumn{1}{c|}{$\rho_c$(mol$\cdot$m$^{-3})$} & \multicolumn{1}{c|}{$P_c$(bar)} \\
\hline
Kirshenbaum \ea\cite{Kirshenbaum1962}/G\cite{Grosse1961}  & $3400 (3200,3600)$ & $4313 (3457,4456)$& $355(228,482)$ \\
Kirshenbaum \ea\cite{Kirshenbaum1962} $ [1738,3200]$/RF\cite{Rodrigues2007}  & $3393 (3285,3750)$ & $4481(4362,4920)$ & $-$
\\
NJ $[1200,2100]$/RF\cite{Rodrigues2007}  & $3225(3189,3264)$ & $4283(4263,4303)$ & $248(231,267)$ \\
JC $[1200,1700]$/RF\cite{Rodrigues2007} & 2601(2590,2612) & $6381(6294,6468)$ & $537(519,555$)\\
MFC-MSC $[1200,2000]$/RF\cite{Rodrigues2007}& $2539(2503,2580)$ & $1865(1732,1983)$ & $150(124,186)$\\
\hline
\end{tabular}
\caption{NaCl critical properties calculated using the indicated data sources and analysis methods.  The intervals in the first column denote the temperature ranges of VLE data used, and in the other columns they indicate uncertainty intervals at the 95\% confidence level; details are given  in Section \ref{sec:UncertaintyCritical}.
}
\label{table:critical}
\end{center}
\end{table}

Kirshenbaum \ea\cite{Kirshenbaum1962} obtained the NaCl critical properties by using a method\cite{Grosse1961} of extrapolating the liquid-phase density curve at temperatures at and above its normal boiling point (nbp) of 1073 K to its intersection with the RLD line calculated from lower temperature (1073 K to 1738 K) VLE data.  At higher temperatures, experimental VLE vapor pressure data were used to compute liquid densities consistent with the extrapolated RLD line.  The resulting high-temperature VLE data set is given in Table \ref{table:PstarK}.  The resulting critical properties are given ini the first row of Table \ref{table:critical}.

We first re-analyzed the Kirshenbaum \ea\cite{Kirshenbaum1962} VLE data set using the scaling-law-based  methodology of Rodrigues and Fernandes\cite{Rodrigues2007}. The results in the second row of Table \ref{table:critical} (regression coefficients  $(E_0,E_1,F_0,F_1)$ corresponding to each of the last 4 rows are given in the SI) indicate that the uncertainty intervals overlap those of Kirshenbaum \ea\cite{Kirshenbaum1962} in the first row, and show a considerably smaller $\rho_c$ uncertainty range. $P_c$ in the second row is blank because Kirshenbaum\ea did not provide $P^*(T)$ in their paper.

To investigate the effect of using only low-temperature VLE data where the vapor phase exhibits ideal-gas behavior, we repeated the Rodrigues and Fernandes\cite{Rodrigues2007}  analysis  using our experimentally-based NJ VLE data at $P^*<10$ bar from Table \ref{table:Pstarvalues}. The results in the third  row of Table \ref{table:critical} indicate that the critical property uncertainty intervals overlap the (relatively large) uncertainty intervals of Kirshenbaum \ea\cite{Kirshenbaum1962} in the first row.  However, they lie slightly lower than those of  the second row.

Finally, the last  two rows of Table \ref{table:critical}, based on results from the framework of Section \ref{sec:framework1} using the simulation data of Kussainova \ea\cite{Kussainova2020}, both give $T_c$ values well below those of the experimentally-based results of the first two rows. The JC value of $\rho_c$ is very high and the MFC-MSC value is very low.

The VLE coexistence curves are shown in Fig. \ref{fig:dome}.  The Joung-Cheatham force field gives better overall results than the hybrid  MFC-MSC methodology.

\subsection{Uncertainty analysis}\label{sec:uncertainty}
\subsubsection{Vapor Pressure}\label{sec:uncertaintyP}
The uncertainties $\Delta \ln(P^*/P^0)$ in $\ln(P^*)$ calculated from Eqs. (\ref{eq:Pstarmonomer2})-(\ref{eq:Pfinal3}) arise from  uncertainties in the input quantities \\ \mbox{$\{x_i\}= \{\Delta G_{r1}(T,P^0)/RT, \Delta G_{r2 }(T,P^0)/RT, \mu^{{\rm res},NPT}_{\rm NaCl,liq}(T,P^0)/RT, \rho_{\rm liq}(T,P^0)\}$}.   The sensitivity coefficients $\partial \ln(P^*/P^0)/\partial x_i$  are derived in  the SI.   We use the NIST-JANAF thermochemical data for $\Delta G^0_{r1}(T,P^0)$ and $\Delta G^0_{r2}(T,P^0)$ and consider these quantities to be exact.  The density sensitivity coefficient is very small and its contribution to the uncertainty may be neglected, and the sensitivity coefficient with respect to $\mu^{{\rm res},\;NPT}(T,P^0)$ gives an uncertainty in $\ln P^*$ of
\begin{equation}
\Delta \ln [P^*] = \frac{\partial \ln P^*}{\partial (\mu^{{\rm res},NPT}/RT)}\Delta (\mu^{{\rm res},\;NPT}(T,P^0)/RT)=(2-y)\Delta (\mu^{{\rm res},\;NPT}(T,P^0)/RT)
 \label{eq:PstarUncertainty}
\end{equation}
We use chemical potential simulation standard deviations reported by Kussainova \ea\cite{Kussainova2020}, and our calculated values of $y$ at each temperature.  We find that the JC and MFC-MSC $P^*$ values vary slightly with $T$ and are approximately 6\% and 18\%, respectively.  They are indicated by the error bars in Fig. \ref{fig:ffcomparison}.
\subsubsection{Vaporization Enthalpy}\label{sec:uncertaintyH}

For the uncertainties of $\Delta H^{\rm vap}_{\rm AB}(T,P^*)$ of Eq. (\ref{eq:deltaHvapfinz}), we used (see  the SI for the derivation)
\begin{equation}
\Delta \Delta H^{\rm AB}_{\rm vap}(T,P^*) =({\bf s(T)}^T{\bf V} {\bf s(T)})^{1/2}
\end{equation}
where {\bf V} is the variance-covariance matrix of the $\mu^{{\rm res},\;NVT}(T,P^0)/T$ regression of Eq. (\ref{eq:Gfit}),
\begin{eqnarray}
{\bf s}(T) & = & \left(
\begin{array}{c}
\omega\\
\omega/T-(2-y)\\
\omega\ln(T)+(2-y)T
\end{array}
\right)
\end{eqnarray}
and
\begin{equation}
\omega = \frac{y(1-y)[\Delta H^0_{\rm r2}(T,P^0)-\Delta H^0_{r1}(T,P^0)-h^{{\rm res},\;NPT}(T,P^0)]}{R}
\end{equation}

We found that the JC and MFC-MSC $\Delta H^{\rm vap}_{\rm AB}(T,P^*)$ values have uncertainties (one standard deviation) of approximately 0.9 kJ$\cdot$mol$^{-1}$ and 4.5 kJ$\cdot$mol$^{-1}$, respectively.  These are indicated as error bars in Fig. \ref{fig:enthalpy}
\subsubsection{Critical Properties}\label{sec:UncertaintyCritical}

An  approximate uncertainty interval at the 95\% confidence level for the $T_c$ nonlinear regression of Eq. (\ref{eq:regress1})  can be determined by calculating the upper ($T_{\rm upper}$) and lower ($T_{\rm lower}$) $T$ values that satisfy\cite{Beale1960}
\begin{equation}
S(T)\leq S(T_c)\left(1+\frac{m}{n-m} F_{0.05}(m,n-m)\right) \label{eq:Fdist}
\end{equation}
where $S$ is the regression's residual sum of squares, $S(T_c)$ is its minimum at the calculated $T_c$ value, $n$ is the number of data points, $m$ is the number of parameters, and $F_{0.05}(m,n-m)$ is the value of the $F$ distribution at the $\alpha= 0.05$ level.  The calculated  $T_{\rm upper}$ and $T_{\rm lower}$  values satisfying Eq. (\ref{eq:Fdist}) were used to determine the uncertainty intervals given in Table \ref{table:critical}.

Following the calculation of the $T_c$ uncertainty, we performed linear regressions of Eq. (\ref{eq:regress2}) at each of $T_{\rm lower}, T_{\rm upper}, T_c$ to determine the upper and lower 95\% confidence limits for the parameter $\rho_c$ in each case.  The $\rho_c$ uncertainty intervals in Table \ref{table:critical} indicate upper and lower extremes of these limits.
\subsection{Conclusions}

We have presented a combined thermodynamic-atomistic simulation framework to calculate the vapor-liquid equilibrium (VLE) properties of a molten salt at temperatures for which the vapor pressure $P^*(T)$ is low (less than 10 bar) and shown its results for NaCl.  It is  based on the assumption that the vapor phase exhibits ideal-gas behavior, which is an excellent approximation for molten salts over a salt-specific low-temperature range.

The framework provides methodology to calculate the temperature dependencies of vapor pressure, coexisting liquid and vapor-phase densities,  vaporization enthalpy, and vapor-phase species concentrations (including those of very small concentration). We also show how to use the liquid and vapor coexistence densities to calculate approximations to the entire VLE coexistence curve and the critical properties.  We  provide uncertainty estimates for the vapor pressure, vaporization enthalpy, and critical properties.

The framework requires knowledge of  the temperature dependence of ideal-gas standard Gibbs energy changes for the vapor-phase reactions (in the case of NaCl, these are the  monomer ionization and dimerization reactions), and data for the liquid densities and residual chemical potentials. If  the ideal-gas data is not available, it can be calculated using electronic structure software such as Gaussian\cite{Gaussian16}, and the liquid chemical potential and density data can be calculated by classical force-field-based simulation methodology.

We validated the framework for NaCl by using experimentally based data from standard sources.  We found that the assumption of only monomer and dimer species in the vapor phase is sufficient to provide results in excellent agreement with the experimental vapor phase measurements up to 2100 K. alculations are given for additional vapor-phase species present in very small concentrations; for example, at 1200 K and 2100 K, the Na$^+$ and Cl$^-$ ion mole fractions are respectively the order of $10^{-9}$  and 10$^{-6}$.   We also use the calculated VLE $T$ and coexisting phase densities to determine the VLE coexistence curve and the molten salt's critical properties, using the scaling law methodology of Rodrigues and Fernandes\cite{Rodrigues2007}.  We found that the use of only low-temperature experimentally based VLE data where the vapor phase exhibits ideal-gas behavior gives slightly low values of the critical properties.

We then applied the framework to NaCl simulation data from the literature\cite{Kussainova2020} for liquid chemical potential and density data determined from the Joung-Cheatham force field\cite{Joung2008} and from the Madrid force field of Benavides \ea\cite{Benavides2017}.  We found that these force fields are inadequate to provide accurate predictions of the NaCl  VLE properties, although the former provides good estimates of the vaporization enthalpy.  This means that molten salt force fields whose parameters are tuned to ambient conditions (such as those considered here) may not be suitable for accurate predictions of high-temperature VLE data.

The most important aspect of the framework is its thermodynamically rigorous incorporation of the vapor-phase species and their chemical reactions.   For NaCl, the most important of these are the monomer ionization and dimerization reactions.   The latter reaction was not considered in previous VLE simulation  studies\cite{Guissani1994,Rodrigues2007,Kussainova2020}.

Provided that the identities of the vapor-phase species are known or can be determined, the framework of this paper may be used to calculate the VLE properties of other molten salts, and of room-temperature ionic liquids.

\clearpage
\noindent {\bf SUPPLEMENTARY MATERIAL\\}
The Supplementary Material contains uncertainty analysis derivations and data tables.\\

\noindent {\bf ACKNOWLEDGMENTS\\}
Financial support was provided by the Natural Science and Engineering Council of Canada (NSERC) and the Agence Nationale de la Recherche (ANR) through the International Collaborative Strategic Program between Canada and France (grants NSERC STPGP 479466-15 and ANR-12-IS09-0001-01). We thank our industrial partner, Dr. John Carroll, Gas Liquids Engineering Ltd., for supporting this research and for helpful advice and encouragement. Computational facilities of the SHARCNET (Shared Hierarchical Academic Research
Computing Network) HPC consortium (www.sharcnet.ca) and Compute Canada (www.computecanada.ca) are gratefully acknowledged.  The authors thank Professor Athanassios Z. Panagiotopoulos for kindly providing numerical data for figures in reference [20].

\clearpage

\begin{thebibliography}{43}%
\makeatletter
\providecommand \@ifxundefined [1]{%
 \@ifx{#1\undefined}
}%
\providecommand \@ifnum [1]{%
 \ifnum #1\expandafter \@firstoftwo
 \else \expandafter \@secondoftwo
 \fi
}%
\providecommand \@ifx [1]{%
 \ifx #1\expandafter \@firstoftwo
 \else \expandafter \@secondoftwo
 \fi
}%
\providecommand \natexlab [1]{#1}%
\providecommand \enquote  [1]{``#1''}%
\providecommand \bibnamefont  [1]{#1}%
\providecommand \bibfnamefont [1]{#1}%
\providecommand \citenamefont [1]{#1}%
\providecommand \href@noop [0]{\@secondoftwo}%
\providecommand \href [0]{\begingroup \@sanitize@url \@href}%
\providecommand \@href[1]{\@@startlink{#1}\@@href}%
\providecommand \@@href[1]{\endgroup#1\@@endlink}%
\providecommand \@sanitize@url [0]{\catcode `\\12\catcode `\$12\catcode
  `\&12\catcode `\#12\catcode `\^12\catcode `\_12\catcode `\%12\relax}%
\providecommand \@@startlink[1]{}%
\providecommand \@@endlink[0]{}%
\providecommand \url  [0]{\begingroup\@sanitize@url \@url }%
\providecommand \@url [1]{\endgroup\@href {#1}{\urlprefix }}%
\providecommand \urlprefix  [0]{URL }%
\providecommand \Eprint [0]{\href }%
\providecommand \doibase [0]{https://doi.org/}%
\providecommand \selectlanguage [0]{\@gobble}%
\providecommand \bibinfo  [0]{\@secondoftwo}%
\providecommand \bibfield  [0]{\@secondoftwo}%
\providecommand \translation [1]{[#1]}%
\providecommand \BibitemOpen [0]{}%
\providecommand \bibitemStop [0]{}%
\providecommand \bibitemNoStop [0]{.\EOS\space}%
\providecommand \EOS [0]{\spacefactor3000\relax}%
\providecommand \BibitemShut  [1]{\csname bibitem#1\endcsname}%
\let\auto@bib@innerbib\@empty
\bibitem [{\citenamefont {Le~Brun}(2007)}]{LeBrun2007}%
  \BibitemOpen
  \bibfield  {author} {\bibinfo {author} {\bibfnamefont {C.}~\bibnamefont
  {Le~Brun}},\ }\href@noop {} {\bibfield  {journal} {\bibinfo  {journal} {J.
  Nuclear Materials}\ }\textbf {\bibinfo {volume} {360}},\ \bibinfo {pages} {1}
  (\bibinfo {year} {2007})}\BibitemShut {NoStop}%
\bibitem [{\citenamefont {Benes}\ and\ \citenamefont
  {Konings}(2009)}]{Benes2009}%
  \BibitemOpen
  \bibfield  {author} {\bibinfo {author} {\bibfnamefont {O.}~\bibnamefont
  {Benes}}\ and\ \bibinfo {author} {\bibfnamefont {R.~J.~M.}\ \bibnamefont
  {Konings}},\ }\href@noop {} {\bibfield  {journal} {\bibinfo  {journal}
  {Journal of Fluorine Chemistry}\ }\textbf {\bibinfo {volume} {130}},\
  \bibinfo {pages} {22} (\bibinfo {year} {2009})}\BibitemShut {NoStop}%
\bibitem [{\citenamefont {Ladkany}\ \emph {et~al.}(2018)\citenamefont
  {Ladkany}, \citenamefont {Culbreth},\ and\ \citenamefont
  {Loyd}}]{Ladkany2018}%
  \BibitemOpen
  \bibfield  {author} {\bibinfo {author} {\bibfnamefont {S.}~\bibnamefont
  {Ladkany}}, \bibinfo {author} {\bibfnamefont {W.}~\bibnamefont {Culbreth}},\
  and\ \bibinfo {author} {\bibfnamefont {N.}~\bibnamefont {Loyd}},\ }\href@noop
  {} {\bibfield  {journal} {\bibinfo  {journal} {J. Energy and Power
  Engineering}\ }\textbf {\bibinfo {volume} {12}},\ \bibinfo {pages} {507}
  (\bibinfo {year} {2018})}\BibitemShut {NoStop}%
\bibitem [{\citenamefont {Li}\ \emph {et~al.}(2021)\citenamefont {Li},
  \citenamefont {K\"{u}\c{c}\"{u}kbenli}, \citenamefont {Lam}, \citenamefont
  {Khaykovich}, \citenamefont {Kaxiras},\ and\ \citenamefont {Li}}]{Li2021}%
  \BibitemOpen
  \bibfield  {author} {\bibinfo {author} {\bibfnamefont {Q.-J.}\ \bibnamefont
  {Li}}, \bibinfo {author} {\bibfnamefont {E.}~\bibnamefont
  {K\"{u}\c{c}\"{u}kbenli}}, \bibinfo {author} {\bibfnamefont {S.}~\bibnamefont
  {Lam}}, \bibinfo {author} {\bibfnamefont {B.}~\bibnamefont {Khaykovich}},
  \bibinfo {author} {\bibfnamefont {E.}~\bibnamefont {Kaxiras}},\ and\ \bibinfo
  {author} {\bibfnamefont {J.}~\bibnamefont {Li}},\ }\href@noop {} {\bibfield
  {journal} {\bibinfo  {journal} {Cell Reports Physical Science}\ }\textbf
  {\bibinfo {volume} {2}} (\bibinfo {year} {2021})}\BibitemShut {NoStop}%
\bibitem [{\citenamefont {Wang}\ \emph {et~al.}(2021)\citenamefont {Wang},
  \citenamefont {Rincon}, \citenamefont {Li}, \citenamefont {Zhao},\ and\
  \citenamefont {Vidal}}]{Wang2021}%
  \BibitemOpen
  \bibfield  {author} {\bibinfo {author} {\bibfnamefont {X.}~\bibnamefont
  {Wang}}, \bibinfo {author} {\bibfnamefont {J.~D.}\ \bibnamefont {Rincon}},
  \bibinfo {author} {\bibfnamefont {P.}~\bibnamefont {Li}}, \bibinfo {author}
  {\bibfnamefont {Y.}~\bibnamefont {Zhao}},\ and\ \bibinfo {author}
  {\bibfnamefont {J.}~\bibnamefont {Vidal}},\ }\href@noop {} {\bibfield
  {journal} {\bibinfo  {journal} {J. Solar Energy Engineering}\ }\textbf
  {\bibinfo {volume} {143}} (\bibinfo {year} {2021})}\BibitemShut {NoStop}%
\bibitem [{\citenamefont {Zhao}()}]{Zhao2021}%
  \BibitemOpen
  \bibfield  {author} {\bibinfo {author} {\bibfnamefont {Y.}~\bibnamefont
  {Zhao}},\ }in\ \href@noop {} {\emph {\bibinfo {booktitle} {Gen3 CSP Summit,
  Golden Co, 25-26 Aug. 2021, NREL/TP-5500-78047}}}\ (\bibinfo  {publisher}
  {NREL/TP-5500-78047})\BibitemShut {NoStop}%
\bibitem [{\citenamefont {Caraballo}\ \emph {et~al.}(2021)\citenamefont
  {Caraballo}, \citenamefont {Galan-Casado}, \citenamefont {Caballero},\ and\
  \citenamefont {Serena}}]{Caraballo2021}%
  \BibitemOpen
  \bibfield  {author} {\bibinfo {author} {\bibfnamefont {A.}~\bibnamefont
  {Caraballo}}, \bibinfo {author} {\bibfnamefont {S.}~\bibnamefont
  {Galan-Casado}}, \bibinfo {author} {\bibfnamefont {A.}~\bibnamefont
  {Caballero}},\ and\ \bibinfo {author} {\bibfnamefont {S.}~\bibnamefont
  {Serena}},\ }\href@noop {} {\bibfield  {journal} {\bibinfo  {journal}
  {Energies}\ }\textbf {\bibinfo {volume} {14}} (\bibinfo {year}
  {2021})}\BibitemShut {NoStop}%
\bibitem [{\citenamefont {Janz}\ \emph {et~al.}(1979)\citenamefont {Janz},
  \citenamefont {Allen}, \citenamefont {Bansal}, \citenamefont {Murphy},\ and\
  \citenamefont {Tomkins}}]{Janz1979}%
  \BibitemOpen
  \bibfield  {author} {\bibinfo {author} {\bibfnamefont {G.}~\bibnamefont
  {Janz}}, \bibinfo {author} {\bibfnamefont {C.}~\bibnamefont {Allen}},
  \bibinfo {author} {\bibfnamefont {N.}~\bibnamefont {Bansal}}, \bibinfo
  {author} {\bibfnamefont {R.}~\bibnamefont {Murphy}},\ and\ \bibinfo {author}
  {\bibfnamefont {R.}~\bibnamefont {Tomkins}},\ }\href@noop {} {\bibfield
  {journal} {\bibinfo  {journal} {NSRDS-NBS}\ }\textbf {\bibinfo {volume} {61,
  Part II}} (\bibinfo {year} {1979})}\BibitemShut {NoStop}%
\bibitem [{\citenamefont {Wang}\ \emph {et~al.}(2020)\citenamefont {Wang},
  \citenamefont {DeFever}, \citenamefont {Zhang}, \citenamefont {Wu},
  \citenamefont {Roy}, \citenamefont {Bryantsev}, \citenamefont {Margulis},\
  and\ \citenamefont {Maginn}}]{Wang2020}%
  \BibitemOpen
  \bibfield  {author} {\bibinfo {author} {\bibfnamefont {H.}~\bibnamefont
  {Wang}}, \bibinfo {author} {\bibfnamefont {R.~S.}\ \bibnamefont {DeFever}},
  \bibinfo {author} {\bibfnamefont {Y.}~\bibnamefont {Zhang}}, \bibinfo
  {author} {\bibfnamefont {F.}~\bibnamefont {Wu}}, \bibinfo {author}
  {\bibfnamefont {S.}~\bibnamefont {Roy}}, \bibinfo {author} {\bibfnamefont
  {V.~S.}\ \bibnamefont {Bryantsev}}, \bibinfo {author} {\bibfnamefont {C.~J.}\
  \bibnamefont {Margulis}},\ and\ \bibinfo {author} {\bibfnamefont {E.~J.}\
  \bibnamefont {Maginn}},\ }\href@noop {} {\bibfield  {journal} {\bibinfo
  {journal} {J. Chem. Phys.}\ }\textbf {\bibinfo {volume} {153}},\ \bibinfo
  {pages} {214502} (\bibinfo {year} {2020})}\BibitemShut {NoStop}%
\bibitem [{\citenamefont {Sharma}\ \emph {et~al.}(2021)\citenamefont {Sharma},
  \citenamefont {Ivanov},\ and\ \citenamefont {Margulis}}]{Sharma2021}%
  \BibitemOpen
  \bibfield  {author} {\bibinfo {author} {\bibfnamefont {S.}~\bibnamefont
  {Sharma}}, \bibinfo {author} {\bibfnamefont {A.~S.}\ \bibnamefont {Ivanov}},\
  and\ \bibinfo {author} {\bibfnamefont {C.~J.}\ \bibnamefont {Margulis}},\
  }\href@noop {} {\bibfield  {journal} {\bibinfo  {journal} {J. Phys. Chem. B}\
  }\textbf {\bibinfo {volume} {125}},\ \bibinfo {pages} {6359} (\bibinfo {year}
  {2021})}\BibitemShut {NoStop}%
\bibitem [{\citenamefont {Panagiotopoulos}(1992)}]{Panagiotopoulos1992}%
  \BibitemOpen
  \bibfield  {author} {\bibinfo {author} {\bibfnamefont {A.~Z.}\ \bibnamefont
  {Panagiotopoulos}},\ }\href@noop {} {\bibfield  {journal} {\bibinfo
  {journal} {Fluid Phase Equilb.}\ }\textbf {\bibinfo {volume} {76}},\ \bibinfo
  {pages} {97} (\bibinfo {year} {1992})}\BibitemShut {NoStop}%
\bibitem [{\citenamefont {Panagiotopoulos}(1987)}]{Panagiotopoulos1987}%
  \BibitemOpen
  \bibfield  {author} {\bibinfo {author} {\bibfnamefont {A.}~\bibnamefont
  {Panagiotopoulos}},\ }\href@noop {} {\bibfield  {journal} {\bibinfo
  {journal} {Molec. Phys.}\ }\textbf {\bibinfo {volume} {61}},\ \bibinfo
  {pages} {813} (\bibinfo {year} {1987})}\BibitemShut {NoStop}%
\bibitem [{\citenamefont {Panagiotopoulos}\ \emph {et~al.}(1988)\citenamefont
  {Panagiotopoulos}, \citenamefont {Quirke}, \citenamefont {Stapleton},\ and\
  \citenamefont {Tildesley}}]{Panagiotopoulos1988}%
  \BibitemOpen
  \bibfield  {author} {\bibinfo {author} {\bibfnamefont {A.~Z.}\ \bibnamefont
  {Panagiotopoulos}}, \bibinfo {author} {\bibfnamefont {N.}~\bibnamefont
  {Quirke}}, \bibinfo {author} {\bibfnamefont {M.}~\bibnamefont {Stapleton}},\
  and\ \bibinfo {author} {\bibfnamefont {D.~J.}\ \bibnamefont {Tildesley}},\
  }\href@noop {} {\bibfield  {journal} {\bibinfo  {journal} {Molec. Phys.}\
  }\textbf {\bibinfo {volume} {64}},\ \bibinfo {pages} {527} (\bibinfo {year}
  {1988})}\BibitemShut {NoStop}%
\bibitem [{\citenamefont {Guissani}\ and\ \citenamefont
  {Guillot}(1994)}]{Guissani1994}%
  \BibitemOpen
  \bibfield  {author} {\bibinfo {author} {\bibfnamefont {Y.}~\bibnamefont
  {Guissani}}\ and\ \bibinfo {author} {\bibfnamefont {B.}~\bibnamefont
  {Guillot}},\ }\href@noop {} {\bibfield  {journal} {\bibinfo  {journal} {J.
  Chem. Phys.}\ }\textbf {\bibinfo {volume} {101}},\ \bibinfo {pages} {490}
  (\bibinfo {year} {1994})}\BibitemShut {NoStop}%
\bibitem [{\citenamefont {Funi}\ and\ \citenamefont {Tosi}(1964)}]{Fumi1964}%
  \BibitemOpen
  \bibfield  {author} {\bibinfo {author} {\bibfnamefont {F.}~\bibnamefont
  {Funi}}\ and\ \bibinfo {author} {\bibfnamefont {M.}~\bibnamefont {Tosi}},\
  }\href@noop {} {\bibfield  {journal} {\bibinfo  {journal} {J. Phys. Chem.
  Solids}\ }\textbf {\bibinfo {volume} {25}},\ \bibinfo {pages} {31} (\bibinfo
  {year} {1964})}\BibitemShut {NoStop}%
\bibitem [{\citenamefont {Tosi}\ and\ \citenamefont {Fumi}(1964)}]{Tosi1964}%
  \BibitemOpen
  \bibfield  {author} {\bibinfo {author} {\bibfnamefont {M.}~\bibnamefont
  {Tosi}}\ and\ \bibinfo {author} {\bibfnamefont {F.}~\bibnamefont {Fumi}},\
  }\href@noop {} {\bibfield  {journal} {\bibinfo  {journal} {J. Phys. Chem.
  Solids}\ }\textbf {\bibinfo {volume} {25}},\ \bibinfo {pages} {45} (\bibinfo
  {year} {1964})}\BibitemShut {NoStop}%
\bibitem [{\citenamefont {Lewis}\ and\ \citenamefont
  {Singer}(1975)}]{Lewis1975}%
  \BibitemOpen
  \bibfield  {author} {\bibinfo {author} {\bibfnamefont {J.}~\bibnamefont
  {Lewis}}\ and\ \bibinfo {author} {\bibfnamefont {K.}~\bibnamefont {Singer}},\
  }\href@noop {} {\bibfield  {journal} {\bibinfo  {journal} {J. Chem. Soc.
  Faraday Trans.}\ }\textbf {\bibinfo {volume} {71}},\ \bibinfo {pages} {41}
  (\bibinfo {year} {1975})}\BibitemShut {NoStop}%
\bibitem [{\citenamefont {Rodrigues}\ and\ \citenamefont
  {Silva~Fernandes}(2007)}]{Rodrigues2007}%
  \BibitemOpen
  \bibfield  {author} {\bibinfo {author} {\bibfnamefont {P.~C.}\ \bibnamefont
  {Rodrigues}}\ and\ \bibinfo {author} {\bibfnamefont {F.~M.}\ \bibnamefont
  {Silva~Fernandes}},\ }\href@noop {} {\bibfield  {journal} {\bibinfo
  {journal} {J Chem Phys}\ }\textbf {\bibinfo {volume} {126}},\ \bibinfo
  {pages} {024503} (\bibinfo {year} {2007})}\BibitemShut {NoStop}%
\bibitem [{\citenamefont {Abramo}\ \emph {et~al.}(2018)\citenamefont {Abramo},
  \citenamefont {Costa}, \citenamefont {Malescio}, \citenamefont {Munao},
  \citenamefont {Pellicane}, \citenamefont {Prestipino},\ and\ \citenamefont
  {Caccamo}}]{Abramo2018}%
  \BibitemOpen
  \bibfield  {author} {\bibinfo {author} {\bibfnamefont {M.~C.}\ \bibnamefont
  {Abramo}}, \bibinfo {author} {\bibfnamefont {D.}~\bibnamefont {Costa}},
  \bibinfo {author} {\bibfnamefont {G.}~\bibnamefont {Malescio}}, \bibinfo
  {author} {\bibfnamefont {G.}~\bibnamefont {Munao}}, \bibinfo {author}
  {\bibfnamefont {G.}~\bibnamefont {Pellicane}}, \bibinfo {author}
  {\bibfnamefont {S.}~\bibnamefont {Prestipino}},\ and\ \bibinfo {author}
  {\bibfnamefont {C.}~\bibnamefont {Caccamo}},\ }\href@noop {} {\bibfield
  {journal} {\bibinfo  {journal} {Phys. Rev. E}\ }\textbf {\bibinfo {volume}
  {98}},\ \bibinfo {pages} {010103} (\bibinfo {year} {2018})}\BibitemShut
  {NoStop}%
\bibitem [{\citenamefont {Kussainova}\ \emph {et~al.}(2020)\citenamefont
  {Kussainova}, \citenamefont {Mondal}, \citenamefont {Young}, \citenamefont
  {Yue},\ and\ \citenamefont {Panagiotopoulos}}]{Kussainova2020}%
  \BibitemOpen
  \bibfield  {author} {\bibinfo {author} {\bibfnamefont {D.}~\bibnamefont
  {Kussainova}}, \bibinfo {author} {\bibfnamefont {A.}~\bibnamefont {Mondal}},
  \bibinfo {author} {\bibfnamefont {J.~M.}\ \bibnamefont {Young}}, \bibinfo
  {author} {\bibfnamefont {S.}~\bibnamefont {Yue}},\ and\ \bibinfo {author}
  {\bibfnamefont {A.~Z.}\ \bibnamefont {Panagiotopoulos}},\ }\href@noop {}
  {\bibfield  {journal} {\bibinfo  {journal} {J. Chem. Phys}\ }\textbf
  {\bibinfo {volume} {153}} (\bibinfo {year} {2020})}\BibitemShut {NoStop}%
\bibitem [{\citenamefont {Joung}\ and\ \citenamefont
  {Cheatham}(2008)}]{Joung2008}%
  \BibitemOpen
  \bibfield  {author} {\bibinfo {author} {\bibfnamefont {I.~S.}\ \bibnamefont
  {Joung}}\ and\ \bibinfo {author} {\bibfnamefont {T.~E.}\ \bibnamefont
  {Cheatham}},\ }\href@noop {} {\bibfield  {journal} {\bibinfo  {journal} {Am.
  Chem. Soc.}\ }\textbf {\bibinfo {volume} {112(30)}},\ \bibinfo {pages} {9021}
  (\bibinfo {year} {2008})}\BibitemShut {NoStop}%
\bibitem [{\citenamefont {Benavides}\ \emph {et~al.}(2017)\citenamefont
  {Benavides}, \citenamefont {Portillo}, \citenamefont {Chamorro},
  \citenamefont {Espinosa}, \citenamefont {Abascal},\ and\ \citenamefont
  {Vega}}]{Benavides2017}%
  \BibitemOpen
  \bibfield  {author} {\bibinfo {author} {\bibfnamefont {A.~L.}\ \bibnamefont
  {Benavides}}, \bibinfo {author} {\bibfnamefont {M.~A.}\ \bibnamefont
  {Portillo}}, \bibinfo {author} {\bibfnamefont {V.~C.}\ \bibnamefont
  {Chamorro}}, \bibinfo {author} {\bibfnamefont {J.~R.}\ \bibnamefont
  {Espinosa}}, \bibinfo {author} {\bibfnamefont {J.~L.~F.}\ \bibnamefont
  {Abascal}},\ and\ \bibinfo {author} {\bibfnamefont {C.}~\bibnamefont
  {Vega}},\ }\href@noop {} {\bibfield  {journal} {\bibinfo  {journal} {J. Chem.
  Phys}\ }\textbf {\bibinfo {volume} {147}},\ \bibinfo {pages} {104501}
  (\bibinfo {year} {2017})}\BibitemShut {NoStop}%
\bibitem [{\citenamefont {Barton}\ and\ \citenamefont
  {Bloom}(1956)}]{Barton1956}%
  \BibitemOpen
  \bibfield  {author} {\bibinfo {author} {\bibfnamefont {J.~L.}\ \bibnamefont
  {Barton}}\ and\ \bibinfo {author} {\bibfnamefont {H.}~\bibnamefont {Bloom}},\
  }\href@noop {} {\bibfield  {journal} {\bibinfo  {journal} {J. Phys. Chem.}\
  }\textbf {\bibinfo {volume} {60}},\ \bibinfo {pages} {1413–1416} (\bibinfo
  {year} {1956})}\BibitemShut {NoStop}%
\bibitem [{\citenamefont {Miller}\ and\ \citenamefont
  {Kusch}(1956)}]{Miller1956}%
  \BibitemOpen
  \bibfield  {author} {\bibinfo {author} {\bibfnamefont {R.~C.}\ \bibnamefont
  {Miller}}\ and\ \bibinfo {author} {\bibfnamefont {P.}~\bibnamefont {Kusch}},\
  }\href@noop {} {\bibfield  {journal} {\bibinfo  {journal} {J. Chem. Phys.}\
  }\textbf {\bibinfo {volume} {25}},\ \bibinfo {pages} {860} (\bibinfo {year}
  {1956})}\BibitemShut {NoStop}%
\bibitem [{\citenamefont {Datz}\ \emph {et~al.}(1961)\citenamefont {Datz},
  \citenamefont {Smith},\ and\ \citenamefont {Taylor}}]{Datz1961}%
  \BibitemOpen
  \bibfield  {author} {\bibinfo {author} {\bibfnamefont {S.}~\bibnamefont
  {Datz}}, \bibinfo {author} {\bibfnamefont {W.~T.}\ \bibnamefont {Smith}},\
  and\ \bibinfo {author} {\bibfnamefont {E.~H.}\ \bibnamefont {Taylor}},\
  }\href@noop {} {\bibfield  {journal} {\bibinfo  {journal} {J. Chem. Phys.}\
  }\textbf {\bibinfo {volume} {34}},\ \bibinfo {pages} {558} (\bibinfo {year}
  {1961})}\BibitemShut {NoStop}%
\bibitem [{\citenamefont {Kvande}(1979)}]{Kvande1979}%
  \BibitemOpen
  \bibfield  {author} {\bibinfo {author} {\bibfnamefont {H.}~\bibnamefont
  {Kvande}},\ }\href@noop {} {\bibfield  {journal} {\bibinfo  {journal} {Acta
  Chemical Scand. A}\ }\textbf {\bibinfo {volume} {33}},\ \bibinfo {pages}
  {407} (\bibinfo {year} {1979})}\BibitemShut {NoStop}%
\bibitem [{\citenamefont {Kvande}\ \emph {et~al.}(1979)\citenamefont {Kvande},
  \citenamefont {Linga}, \citenamefont {Motzfeldt},\ and\ \citenamefont
  {Wahlbeck}}]{Kvande1979b}%
  \BibitemOpen
  \bibfield  {author} {\bibinfo {author} {\bibfnamefont {H.}~\bibnamefont
  {Kvande}}, \bibinfo {author} {\bibfnamefont {H.}~\bibnamefont {Linga}},
  \bibinfo {author} {\bibfnamefont {K.}~\bibnamefont {Motzfeldt}},\ and\
  \bibinfo {author} {\bibfnamefont {P.~G.}\ \bibnamefont {Wahlbeck}},\
  }\href@noop {} {\bibfield  {journal} {\bibinfo  {journal} {Acta Chem. Scand.
  A}\ }\textbf {\bibinfo {volume} {33}},\ \bibinfo {pages} {281} (\bibinfo
  {year} {1979})}\BibitemShut {NoStop}%
\bibitem [{\citenamefont {Pitzer}(1996)}]{Pitzer1996}%
  \BibitemOpen
  \bibfield  {author} {\bibinfo {author} {\bibfnamefont {K.~S.}\ \bibnamefont
  {Pitzer}},\ }\href@noop {} {\bibfield  {journal} {\bibinfo  {journal} {J.
  Chem. Phys.}\ }\textbf {\bibinfo {volume} {14(17)}},\ \bibinfo {pages} {6724}
  (\bibinfo {year} {1996})}\BibitemShut {NoStop}%
\bibitem [{\citenamefont {Chase}(1998)}]{Chase1998}%
  \BibitemOpen
  \bibfield  {author} {\bibinfo {author} {\bibfnamefont {M.~W.~J.}\
  \bibnamefont {Chase}},\ }\href@noop {} {\bibfield  {journal} {\bibinfo
  {journal} {Am. Chem. Society and Am. Physical Society}\ }\textbf {\bibinfo
  {volume} {79}},\ \bibinfo {pages} {102} (\bibinfo {year} {1998})}\BibitemShut
  {NoStop}%
\bibitem [{\citenamefont {Frisch}\ \emph {et~al.}(2016)\citenamefont {Frisch},
  \citenamefont {Trucks}, \citenamefont {Schlegel}, \citenamefont {Scuseria},
  \citenamefont {Robb}, \citenamefont {Cheeseman}, \citenamefont {Scalmani},
  \citenamefont {Barone}, \citenamefont {Petersson}, \citenamefont {Nakatsuji},
  \citenamefont {Li}, \citenamefont {Caricato}, \citenamefont {Marenich},
  \citenamefont {Bloino}, \citenamefont {Janesko}, \citenamefont {Gomperts},
  \citenamefont {Mennucci}, \citenamefont {Hratchian}, \citenamefont {Ortiz},
  \citenamefont {Izmaylov}, \citenamefont {Sonnenberg}, \citenamefont
  {Williams-Young}, \citenamefont {Ding}, \citenamefont {Lipparini},
  \citenamefont {Egidi}, \citenamefont {Goings}, \citenamefont {Peng},
  \citenamefont {Petrone}, \citenamefont {Henderson}, \citenamefont
  {Ranasinghe}, \citenamefont {Zakrzewski}, \citenamefont {Gao}, \citenamefont
  {Rega}, \citenamefont {Zheng}, \citenamefont {Liang}, \citenamefont {Hada},
  \citenamefont {Ehara}, \citenamefont {Toyota}, \citenamefont {Fukuda},
  \citenamefont {Hasegawa}, \citenamefont {Ishida}, \citenamefont {Nakajima},
  \citenamefont {Honda}, \citenamefont {Kitao}, \citenamefont {Naka},
  \citenamefont {Vreven}, \citenamefont {Throssell}, \citenamefont
  {J.~A.~Montgomery}, \citenamefont {Peralta}, \citenamefont {Ogliaro},
  \citenamefont {Bearpark}, \citenamefont {Heyd}, \citenamefont {Brothers},
  \citenamefont {Kudin}, \citenamefont {Staroverov}, \citenamefont {Keith},
  \citenamefont {Kobayashi}, \citenamefont {Normand}, \citenamefont
  {Raghavachari}, \citenamefont {Rendell}, \citenamefont {Burant},
  \citenamefont {Iyengar}, \citenamefont {Tomasi}, \citenamefont {Cossi},
  \citenamefont {Millam}, \citenamefont {Klene}, \citenamefont {Adamo},
  \citenamefont {Cammi}, \citenamefont {Ochterski}, \citenamefont {Martin},
  \citenamefont {Morokuma}, \citenamefont {Farkas}, \citenamefont {Foresman},\
  and\ \citenamefont {Fox}}]{Gaussian16}%
  \BibitemOpen
  \bibfield  {author} {\bibinfo {author} {\bibfnamefont {M.~J.}\ \bibnamefont
  {Frisch}}, \bibinfo {author} {\bibfnamefont {G.~W.}\ \bibnamefont {Trucks}},
  \bibinfo {author} {\bibfnamefont {H.~B.}\ \bibnamefont {Schlegel}}, \bibinfo
  {author} {\bibfnamefont {G.~E.}\ \bibnamefont {Scuseria}}, \bibinfo {author}
  {\bibfnamefont {M.~A.}\ \bibnamefont {Robb}}, \bibinfo {author}
  {\bibfnamefont {J.~R.}\ \bibnamefont {Cheeseman}}, \bibinfo {author}
  {\bibfnamefont {G.}~\bibnamefont {Scalmani}}, \bibinfo {author}
  {\bibfnamefont {V.}~\bibnamefont {Barone}}, \bibinfo {author} {\bibfnamefont
  {G.~A.}\ \bibnamefont {Petersson}}, \bibinfo {author} {\bibfnamefont
  {H.}~\bibnamefont {Nakatsuji}}, \bibinfo {author} {\bibfnamefont
  {X.}~\bibnamefont {Li}}, \bibinfo {author} {\bibfnamefont {M.}~\bibnamefont
  {Caricato}}, \bibinfo {author} {\bibfnamefont {A.~V.}\ \bibnamefont
  {Marenich}}, \bibinfo {author} {\bibfnamefont {J.}~\bibnamefont {Bloino}},
  \bibinfo {author} {\bibfnamefont {B.~G.}\ \bibnamefont {Janesko}}, \bibinfo
  {author} {\bibfnamefont {R.}~\bibnamefont {Gomperts}}, \bibinfo {author}
  {\bibfnamefont {B.}~\bibnamefont {Mennucci}}, \bibinfo {author}
  {\bibfnamefont {H.~P.}\ \bibnamefont {Hratchian}}, \bibinfo {author}
  {\bibfnamefont {J.~V.}\ \bibnamefont {Ortiz}}, \bibinfo {author}
  {\bibfnamefont {A.~F.}\ \bibnamefont {Izmaylov}}, \bibinfo {author}
  {\bibfnamefont {J.~L.}\ \bibnamefont {Sonnenberg}}, \bibinfo {author}
  {\bibfnamefont {D.}~\bibnamefont {Williams-Young}}, \bibinfo {author}
  {\bibfnamefont {F.}~\bibnamefont {Ding}}, \bibinfo {author} {\bibfnamefont
  {F.}~\bibnamefont {Lipparini}}, \bibinfo {author} {\bibfnamefont
  {F.}~\bibnamefont {Egidi}}, \bibinfo {author} {\bibfnamefont
  {J.}~\bibnamefont {Goings}}, \bibinfo {author} {\bibfnamefont
  {B.}~\bibnamefont {Peng}}, \bibinfo {author} {\bibfnamefont {A.}~\bibnamefont
  {Petrone}}, \bibinfo {author} {\bibfnamefont {T.}~\bibnamefont {Henderson}},
  \bibinfo {author} {\bibfnamefont {D.}~\bibnamefont {Ranasinghe}}, \bibinfo
  {author} {\bibfnamefont {V.~G.}\ \bibnamefont {Zakrzewski}}, \bibinfo
  {author} {\bibfnamefont {J.}~\bibnamefont {Gao}}, \bibinfo {author}
  {\bibfnamefont {N.}~\bibnamefont {Rega}}, \bibinfo {author} {\bibfnamefont
  {G.}~\bibnamefont {Zheng}}, \bibinfo {author} {\bibfnamefont
  {W.}~\bibnamefont {Liang}}, \bibinfo {author} {\bibfnamefont
  {M.}~\bibnamefont {Hada}}, \bibinfo {author} {\bibfnamefont {M.}~\bibnamefont
  {Ehara}}, \bibinfo {author} {\bibfnamefont {K.}~\bibnamefont {Toyota}},
  \bibinfo {author} {\bibfnamefont {R.}~\bibnamefont {Fukuda}}, \bibinfo
  {author} {\bibfnamefont {J.}~\bibnamefont {Hasegawa}}, \bibinfo {author}
  {\bibfnamefont {M.}~\bibnamefont {Ishida}}, \bibinfo {author} {\bibfnamefont
  {T.}~\bibnamefont {Nakajima}}, \bibinfo {author} {\bibfnamefont
  {Y.}~\bibnamefont {Honda}}, \bibinfo {author} {\bibfnamefont
  {O.}~\bibnamefont {Kitao}}, \bibinfo {author} {\bibfnamefont
  {H.}~\bibnamefont {Naka}}, \bibinfo {author} {\bibfnamefont {T.}~\bibnamefont
  {Vreven}}, \bibinfo {author} {\bibfnamefont {K.}~\bibnamefont {Throssell}},
  \bibinfo {author} {\bibfnamefont {J.}~\bibnamefont {J.~A.~Montgomery}},
  \bibinfo {author} {\bibfnamefont {J.~E.}\ \bibnamefont {Peralta}}, \bibinfo
  {author} {\bibfnamefont {F.}~\bibnamefont {Ogliaro}}, \bibinfo {author}
  {\bibfnamefont {M.~J.}\ \bibnamefont {Bearpark}}, \bibinfo {author}
  {\bibfnamefont {J.~J.}\ \bibnamefont {Heyd}}, \bibinfo {author}
  {\bibfnamefont {E.~N.}\ \bibnamefont {Brothers}}, \bibinfo {author}
  {\bibfnamefont {K.~N.}\ \bibnamefont {Kudin}}, \bibinfo {author}
  {\bibfnamefont {V.~N.}\ \bibnamefont {Staroverov}}, \bibinfo {author}
  {\bibfnamefont {T.~A.}\ \bibnamefont {Keith}}, \bibinfo {author}
  {\bibfnamefont {R.}~\bibnamefont {Kobayashi}}, \bibinfo {author}
  {\bibfnamefont {J.}~\bibnamefont {Normand}}, \bibinfo {author} {\bibfnamefont
  {K.}~\bibnamefont {Raghavachari}}, \bibinfo {author} {\bibfnamefont {A.~P.}\
  \bibnamefont {Rendell}}, \bibinfo {author} {\bibfnamefont {J.~C.}\
  \bibnamefont {Burant}}, \bibinfo {author} {\bibfnamefont {S.~S.}\
  \bibnamefont {Iyengar}}, \bibinfo {author} {\bibfnamefont {J.}~\bibnamefont
  {Tomasi}}, \bibinfo {author} {\bibfnamefont {M.}~\bibnamefont {Cossi}},
  \bibinfo {author} {\bibfnamefont {J.~M.}\ \bibnamefont {Millam}}, \bibinfo
  {author} {\bibfnamefont {M.}~\bibnamefont {Klene}}, \bibinfo {author}
  {\bibfnamefont {C.}~\bibnamefont {Adamo}}, \bibinfo {author} {\bibfnamefont
  {R.}~\bibnamefont {Cammi}}, \bibinfo {author} {\bibfnamefont {J.~W.}\
  \bibnamefont {Ochterski}}, \bibinfo {author} {\bibfnamefont {R.~L.}\
  \bibnamefont {Martin}}, \bibinfo {author} {\bibfnamefont {K.}~\bibnamefont
  {Morokuma}}, \bibinfo {author} {\bibfnamefont {O.}~\bibnamefont {Farkas}},
  \bibinfo {author} {\bibfnamefont {J.~B.}\ \bibnamefont {Foresman}},\ and\
  \bibinfo {author} {\bibfnamefont {D.~J.}\ \bibnamefont {Fox}},\ }\href@noop
  {} {\emph {\bibinfo {title} {{Gaussian 16 {R}evision {C}.01}}}}\ (\bibinfo
  {publisher} {Gaussian Inc.},\ \bibinfo {address} {Wallingford CT},\ \bibinfo
  {year} {2016})\ \bibinfo {note} {{Gaussian Inc. Wallingford CT}}\BibitemShut
  {NoStop}%
\bibitem [{\citenamefont {Kirshenbaum}\ \emph {et~al.}(1962)\citenamefont
  {Kirshenbaum}, \citenamefont {Cahill}, \citenamefont {McGonigal},\ and\
  \citenamefont {Grosse}}]{Kirshenbaum1962}%
  \BibitemOpen
  \bibfield  {author} {\bibinfo {author} {\bibfnamefont {A.~D.}\ \bibnamefont
  {Kirshenbaum}}, \bibinfo {author} {\bibfnamefont {J.~A.}\ \bibnamefont
  {Cahill}}, \bibinfo {author} {\bibfnamefont {P.~J.}\ \bibnamefont
  {McGonigal}},\ and\ \bibinfo {author} {\bibfnamefont {A.}~\bibnamefont
  {Grosse}},\ }\href@noop {} {\bibfield  {journal} {\bibinfo  {journal} {J.
  Inorg. Nucl. Chem.}\ }\textbf {\bibinfo {volume} {24}},\ \bibinfo {pages}
  {1287} (\bibinfo {year} {1962})}\BibitemShut {NoStop}%
\bibitem [{\citenamefont {Marcus}(2013)}]{Marcus2013}%
  \BibitemOpen
  \bibfield  {author} {\bibinfo {author} {\bibfnamefont {Y.}~\bibnamefont
  {Marcus}},\ }\href@noop {} {\bibfield  {journal} {\bibinfo  {journal} {J.
  Chem. Thermodynamics}\ }\textbf {\bibinfo {volume} {61}},\ \bibinfo {pages}
  {7} (\bibinfo {year} {2013})}\BibitemShut {NoStop}%
\bibitem [{\citenamefont {Smith}\ and\ \citenamefont
  {Missen}(1991)}]{Smith1991b}%
  \BibitemOpen
  \bibfield  {author} {\bibinfo {author} {\bibfnamefont {W.}~\bibnamefont
  {Smith}}\ and\ \bibinfo {author} {\bibfnamefont {R.}~\bibnamefont {Missen}},\
  }\href@noop {} {\emph {\bibinfo {title} {Chemical Reaction Equilibrium
  Analysis: {T}heory and Algorithms}}}\ (\bibinfo  {publisher} {Krieger
  Publishing Co.; Reprint of same title, Wiley-Interscience, 1982},\ \bibinfo
  {address} {Malabar, Florida},\ \bibinfo {year} {1991})\BibitemShut {NoStop}%
\bibitem [{\citenamefont {Leal}\ \emph {et~al.}(2017)\citenamefont {Leal},
  \citenamefont {Kulik}, \citenamefont {Smith},\ and\ \citenamefont
  {Saar}}]{Leal2017}%
  \BibitemOpen
  \bibfield  {author} {\bibinfo {author} {\bibfnamefont {A.~M.~M.}\
  \bibnamefont {Leal}}, \bibinfo {author} {\bibfnamefont {D.~A.}\ \bibnamefont
  {Kulik}}, \bibinfo {author} {\bibfnamefont {W.~R.}\ \bibnamefont {Smith}},\
  and\ \bibinfo {author} {\bibfnamefont {M.~O.}\ \bibnamefont {Saar}},\
  }\href@noop {} {\bibfield  {journal} {\bibinfo  {journal} {Pure Appl. Chem.}\
  }\textbf {\bibinfo {volume} {89}},\ \bibinfo {pages} {597–643} (\bibinfo
  {year} {2017})}\BibitemShut {NoStop}%
\bibitem [{\citenamefont {Horiba}\ and\ \citenamefont
  {Baba}(1928)}]{Horiba1928}%
  \BibitemOpen
  \bibfield  {author} {\bibinfo {author} {\bibfnamefont {S.}~\bibnamefont
  {Horiba}}\ and\ \bibinfo {author} {\bibfnamefont {H.}~\bibnamefont {Baba}},\
  }\href@noop {} {\ \textbf {\bibinfo {volume} {3}},\ \bibinfo {pages} {11}
  (\bibinfo {year} {1928})}\BibitemShut {NoStop}%
\bibitem [{\citenamefont {Fiock}\ and\ \citenamefont
  {Rodebush}(1926)}]{Fiock1926}%
  \BibitemOpen
  \bibfield  {author} {\bibinfo {author} {\bibfnamefont {E.~F.}\ \bibnamefont
  {Fiock}}\ and\ \bibinfo {author} {\bibfnamefont {W.~H.}\ \bibnamefont
  {Rodebush}},\ }\href@noop {} {\bibfield  {journal} {\bibinfo  {journal} {J.
  Am. Chem. Soc.}\ }\textbf {\bibinfo {volume} {48}},\ \bibinfo {pages} {2522}
  (\bibinfo {year} {1926})}\BibitemShut {NoStop}%
\bibitem [{\citenamefont {Stull}(1947)}]{Stull1947a}%
  \BibitemOpen
  \bibfield  {author} {\bibinfo {author} {\bibfnamefont {D.~R.}\ \bibnamefont
  {Stull}},\ }\href@noop {} {\bibfield  {journal} {\bibinfo  {journal} {Ind. \&
  Eng. Chem.}\ }\textbf {\bibinfo {volume} {39}},\ \bibinfo {pages} {517}
  (\bibinfo {year} {1947})}\BibitemShut {NoStop}%
\bibitem [{\citenamefont {Wartenburg}\ and\ \citenamefont
  {Albrecht}(1921)}]{Von1921}%
  \BibitemOpen
  \bibfield  {author} {\bibinfo {author} {\bibfnamefont {H.~V.}\ \bibnamefont
  {Wartenburg}}\ and\ \bibinfo {author} {\bibfnamefont {P.}~\bibnamefont
  {Albrecht}},\ }\href@noop {} {\bibfield  {journal} {\bibinfo  {journal} {Z.
  Elektrochem}\ }\textbf {\bibinfo {volume} {27}},\ \bibinfo {pages} {162}
  (\bibinfo {year} {1921})}\BibitemShut {NoStop}%
\bibitem [{\citenamefont {Ruff}\ and\ \citenamefont {Mugdan}(1921)}]{Ruff1921}%
  \BibitemOpen
  \bibfield  {author} {\bibinfo {author} {\bibfnamefont {O.}~\bibnamefont
  {Ruff}}\ and\ \bibinfo {author} {\bibfnamefont {S.}~\bibnamefont {Mugdan}},\
  }\href@noop {} {\bibfield  {journal} {\bibinfo  {journal} {Z. Anorg. Allgem.
  Chem.}\ }\textbf {\bibinfo {volume} {117}},\ \bibinfo {pages} {147} (\bibinfo
  {year} {1921})}\BibitemShut {NoStop}%
\bibitem [{\citenamefont {Vega}(2015)}]{Vega2015}%
  \BibitemOpen
  \bibfield  {author} {\bibinfo {author} {\bibfnamefont {C.}~\bibnamefont
  {Vega}},\ }\href@noop {} {\bibfield  {journal} {\bibinfo  {journal} {Molec.
  Phys.}\ }\textbf {\bibinfo {volume} {113}},\ \bibinfo {pages} {1145}
  (\bibinfo {year} {2015})}\BibitemShut {NoStop}%
\bibitem [{\citenamefont {Kelley}(1935)}]{Kelley1935}%
  \BibitemOpen
  \bibfield  {author} {\bibinfo {author} {\bibfnamefont {K.}~\bibnamefont
  {Kelley}},\ }\href@noop {} {\bibfield  {journal} {\bibinfo  {journal} {US
  Bureau of Mines Bulletiin}\ }\textbf {\bibinfo {volume} {383}} (\bibinfo
  {year} {1935})}\BibitemShut {NoStop}%
\bibitem [{\citenamefont {Grosse}(1961)}]{Grosse1961}%
  \BibitemOpen
  \bibfield  {author} {\bibinfo {author} {\bibfnamefont {A.}~\bibnamefont
  {Grosse}},\ }\href@noop {} {\bibfield  {journal} {\bibinfo  {journal} {J.
  Inorg. Nucl. Chem.}\ }\textbf {\bibinfo {volume} {22}},\ \bibinfo {pages}
  {23} (\bibinfo {year} {1961})}\BibitemShut {NoStop}%
\bibitem [{\citenamefont {Beale}(1960)}]{Beale1960}%
  \BibitemOpen
  \bibfield  {author} {\bibinfo {author} {\bibfnamefont {E.}~\bibnamefont
  {Beale}},\ }\href@noop {} {\bibfield  {journal} {\bibinfo  {journal} {J. Roy.
  Soc. B}\ }\textbf {\bibinfo {volume} {22}},\ \bibinfo {pages} {41} (\bibinfo
  {year} {1960})}\BibitemShut {NoStop}%
\end{thebibliography}
\clearpage

\end{document}